# Assertion Based Functional Verification of March Algorithm Based MBIST Controller

*A MASTER'S THESIS*

*Submitted*

*in the partial fulfillment of the requirements for the award of the degree*

*Of*

**MASTER OF TECHNOLOGY**

*In*

**INFORMATION TECHNOLOGY**
**(Specialization in MICROELECTRONICS)**

*Submitted by*

**Ashwani Kumar**

*Under the Guidance of*

**Dr. Kusum Lata**
Faculty (Electronics Dept.)
IIIT-Allahabad

**INDIAN INSTITUTE OF INFORMATION TECHNOLOGY**
**ALLAHABAD – 211012 (INDIA)**

# Chapter 1

# Introduction

### 1.1 Motivation

In the present era, the more data dominating applications on System on Chip (SoCs) require large number of embedded memories. That's why embedded memories are in focus of technology scaling. Due to very small geometries, embedded memories are susceptible to subtle defects [1, 2]. The testing of memories is very crucial because of very low yield of memories during fabrication. But in case of embedded memories it becomes a tedious task because of no primary inputs and outputs. For this purpose a MBIST circuitry is added with memory which make this testing job very comfortable while adding some area over head. Here, MBIST controller is the core of MBIST architecture [3] that controls the events sequencing during memory testing that's why MBIST controller requires more attention towards its design and functional verification.

In complex IC design, verification takes about 70-80% of the design develop time. This portion of development time is still increasing for high speed IC design [4]. Further, in the reuse-based project, main blocks are recycled by using existing designs, making more effort being shifted from the design stage to the verification stage. In this circumstance, to increase the efficiency of verification, reducing the time consumption of verification stage, is therefore crucial to speed up the whole development process. This can be done by parallel design and verification stages in the design development. That's why the verification of MBIST controller is done parallel to its design process. But for functional verification, traditional simulation-based is good only at validating baseline functionality and it has been declared insufficient for detecting critical corner-case errors, referred to herein as verification hot spots[5].

For this reason, an effective method, to increase the observability of the design and to easily find out and diagnose the design flaws, is keenly needed. Assertion-based verification (ABV) is such a method, which combines the formal techniques having assertions and the simulation based traditional function verification. Here code coverage and functional coverage metrics can be used to follow the progress and quality of the verification efforts.

Assertion is used to describe the properties that a design should hold or should never hold. Some properties with time requirements are not easy to be described with Verilog; and usually needs much more lines of codes with VHDL. There are many standardized assertion languages such as PSL, System Verilog Assertion (SVA) and assertion libraries such as OVL, QVL [5] with which the endeavor for describing a design property becomes much easier and efficient. Recently, SVAs as a set of System Verilog language [6], has been more extensively used as specification of assertions within the design, enabling the simulator to check the assertions during simulation.

## 1.2 Analysis of Previous Work in this Area

Now days, embedded memories are dominating to cover the chip area that's why become the focus of technology scaling [1]. Today's data-dominated multimedia applications require more memory than ever before. On-chip SRAM memories begin to dominate the chip area and have become the focus of technology scaling. However, the physical limitations of the technology scaling jeopardize further progress of microelectronics as scaling results in process variations.

In present era, embedded memories cover a large area on the chip for complex designs. So it is one of the most significant step to implement efficient memory testing strategies [2]. MBIST architecture is one solution which automates the memory testing. It also became necessary to scaling down the embedded memories. The small and fine geometries of embedded memories make them susceptible to subtle defects which cause the various faults in memories. That's why, test patterns generation strategies are chosen carefully to detect the manufacturing defects. These algorithms are generally includes the March and checkerboard, varied pattern backgrounds and others.

The MBIST architecture consists of three main blocks controller, pattern generator and signature analyzer [6]. Controller handles the all events in MBIST. It controls such as the increment or decrement in address of memory, whether 0s or 1s patterns is going to read or write at those addresses. The pattern controller block contains the address, data generator and some controls to apply the right pattern to right address. It specifically consists a up and down counter to control the address sequence. Signature analyzer checks whether memory read out data is equal to original generated data or not. Based on checking result the particular flag will be asserted to show pass or fail of pattern.

A FSM MBIST controller is implemented using VHDL and XILINX ISE tool [6]. The FSM defines control signals and shows when one state will proceed to the next state. Each state has its sub-states. The counter is a key element of MBIST architecture because the up and down address sequence is design using counter.

The synthesized design is capable in meeting the functional specifications when implemented in FPGA [6]. For the different patterns, there is a requirement of modification in pattern generator.

The design of data and read/write controller for MARCH algorithm based MBIST architecture to test SRAM [7]. The controller is implemented as an FSM BIST using Verilog HDL to generated test patterns based on MARCH algorithm to detect the SAFs and TFs.

The design is realized by using ALTERA QUARATUS software to generate the RTL abstraction of controller. The SAFs and TFs can be detected and distinguished by generating the different fault syndromes for modified MARCH C algorithm.

$$\{\Updownarrow(w0);\ \Uparrow(r0, w1);\ \Updownarrow(w1, r1);\ \Downarrow(r1, w0, r0);\ \Updownarrow(w0, r0)\}$$

M0 M1 M2 M3 M4

**Figure1.1 March Based Diagnostic Algorithm for SAFs and TFs [7].**

The fault syndrome and indicators are labeled as (R0, R1, R2 …Rn-1) for n read operations during testing. After detecting a fault the read operation generates 1 as fault syndrome, otherwise generates 0. If two faults have same fault syndrome, it means they are detected but they cannot be distinguish. The normal MARCH C algorithm has same fault syndrome for SAFs 0 and TFs 0. A complete state diagram of MBIST read/write is presented for algorithm shown in figure. Controller design can be used to build complete MBIST architecture.

Controller design was also presented with simulation results to show the functionality of the design separately not by considering the issues after its insertion in MBIST architecture.

More number of variations in the fabrication process causes parametric variations in transistor feature sizes and threshold voltages due to random dopant fluctuations, line edge roughness, sub-wavelength lithography [8]. Closely matched devices and small transistor sizes which matter the most when designing SRAM memories, are the first to suffer from the side-effects of scaling. Random nature of local process variation causes defects to have random and uniform distribution .This adversely affects the expected system yield. Since memory is one of the biggest blocks of any system, it is more prone to faults under process variation.

A failure in memory cell can occur due to these following reasons.

a) An increase in cell access time.
b) Unstable read/write operation.
c) Inability to hold the cell content.

The mismatch in device parameters will increase the probability of these failures.

The modeling and simulation of MBIST architecture for embedded memories as a FSM design is presented [9]. Verification of architecture is done by testing stuck at faults in SRAM memory using March C testing algorithm.

The presented MBIST architecture shows the interface connectivity among controller, pattern generator, address generator and pattern/data comparator clearly with interconnect signals [10]. Here, the BIST controller is design for dual port SRAM testing MBIST architecture. BIST controller generates the control signal DGentEN to the data generator to generate data pattern according to March elements in selected march algorithm, control signal AGenEN to address generator to ensure the correct address generation for read/write operation to memory and the control signal DComEN to data comparator. The design architecture of address generator is explained with its functional behavior required for dual port SRAM testing. Design Complier from Synopsys Inc. is used to verify the DPMBIST.

The implementation of formal verification techniques for MBIST controller more specifically, hardwired (as a FSM) memory BIST and programmable (as a micro code control) memory BIST controllers are considered [12].The formal verification of such large controllers is a complex task, than it is tried to simplify this by using the symbolic model checking, verification of MBIST logic, test cases for controller and hookup logic, converting the testcases into formal property and module abstraction.

The conversion of test cases into formal properties of the design helps to validate the test cases which are applied to the design. For example, the validation of the stability condition begins in any State of controller. If the controller is to satisfy stability condition, it means controller would not proceed or acquire the next states. If this state is 'Done' state of MBIST controller then this anticipates that controller should never allow the MBIST_Done signal to be de-asserted. In other words this property shows that FSM of controller should never transit to state where MBIST_Done signal will be de-asserted.

A five-step method is suggested for the functional verification of digital IC verification using System Verilog Assertion (SVAs) [15].

(a) List design interfaces.

(b) List the inside functional spots.

(c) List verification requirement for interface.

(d) Formalize the properties with SVA.

(e) Define coverage spots.

The method is demonstrated for the functional verification of a UART RTL Verilog model. During simulation, the corner cases can be easily checked which are left out in the traditional functional checking. These are corner cases are tested against the SVAs written to check the specific property of the design. This is very much possible that these corner cases consists the bugs in the design but now these design errors can be exposed through the dynamic simulation output (waveform or log file). Furthermore, by this approach, the assertion coverage report can be directly used to validate the design and to show the depth of verification by functional coverage analysis. The suggested five steps increase the observability of design while designing it. This is feasible for being applied in both the design process and verification process of RTL model.

A method which is used to collect the coverage information is shown in Figure1.2 [25]. Coverage information is logged into a database of the tool which is used as a verification tool during simulation of tests with assertions. This data is then parsed using a report generation tool which is called Universal Report Generation (URG), which generates readable reports from the data collected by the tool during the simulation of design.

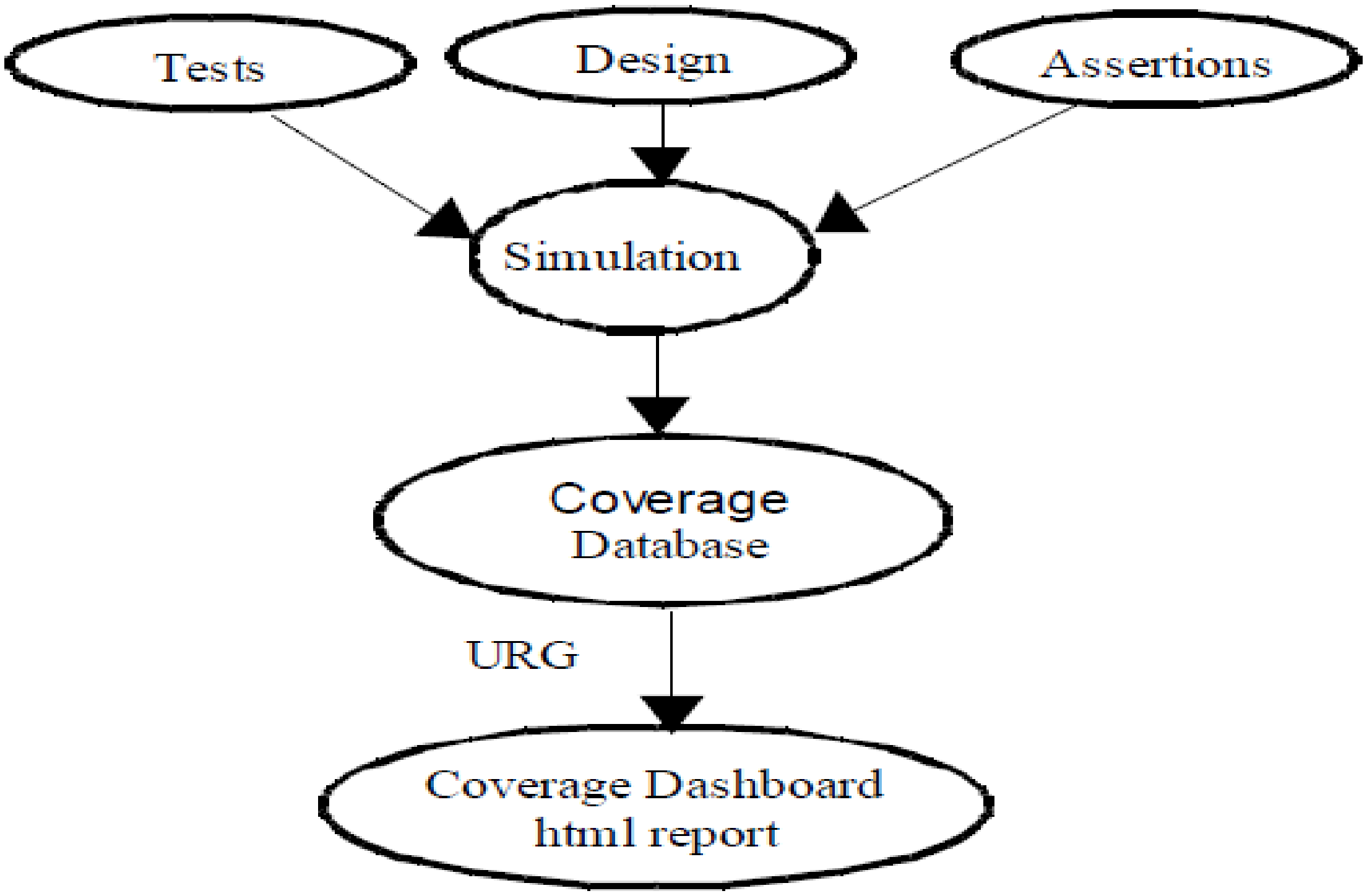

**Figure1.2 Coverage Measurement Flow**

ABV is a necessary part of any design verification process. Assertions allow formal proof with the traditional simulation based proof for the correct design functionality, at the same time it helps to accelerate the bug finding in the design.

Assertions assist in measuring functional coverage but at the same time they increase the cost due to maintenance of overhead and cause degradation in simulation speed. Formal Verification alone cannot be the complete solution for design verification with the existing limitations in design style as it is shown in Figure 1.3 that formal verification has less number of bugs found by using it. But incorporating the assertions into the verification methodology for the CMT processor helped to accelerate the complete design cycle.

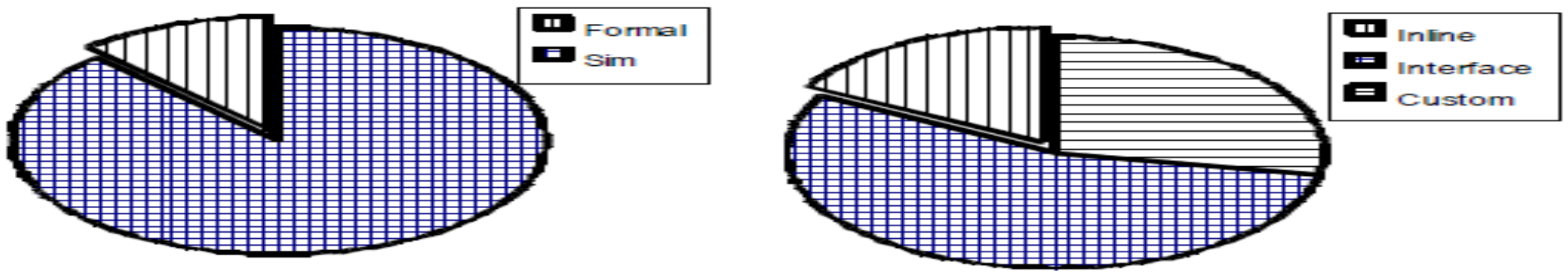

**Figure1.3The distribution of bugs found using assertion checks in formal as well as in simulation check.**

**1.3 Problem Definition and Scope**

The thesis problem is defined as "Assertion Based Functional Verification of March Algorithm Based MBIST Controller".

The parallel verification approach at block level design can be used to implement the complete MBIST architecture to test the single port SRAM.

The presented approach ensures the correct behaviors of the modules after interfacing with other modules.

The March C a memory testing algorithm is also modified to test the memory data retention faults with other memory faults. The design of MBIST controller is also based on this modified March C algorithm.

**1.4 Problem Formulation**

This thesis work is divided in to four main parts.

The first part of the work was to design MBIST controller as a verification module using System Verilog language, for MBIST architecture to test the SRAM.

The second part of the thesis was to learn theory about the verification especially Assertion based dynamic functional verification, verification plan and system Verilog as assertion language.

The third part was to make a verification plan for the MBIST controller's RTL module design. The verification plan of MBIST controller includes the complete specification documentation of MBIST controller, test bench planning, code coverage, functional coverage, assertions etc.

The fourth part was to implement the verification plan. This includes the simulation using Synopsys VCS®, generating the coverage metrics, implementation of only required testcases, analysis of results.

### 1.5 Thesis Formulation

Chapter 2 of thesis is the introductory chapter of MBIST architecture, MBIST controller and memory testing March algorithms. It provides the basic idea behind the designed MBIST controller in this thesis.

Chapter 3 includes the theoretical background of various aspects of verification fields which are used to verify the design in this thesis work. It comprises the verification technologies, verification planning and coverage collection.

Chapter 4 describes the tools and languages briefly which are used to complete the thesis work. The design and verification features are described for Synopsys VCS® as a verification tool and System Verilog as a language for design and verification both.

Chapter 5, 6, 7 and 8 present my contributions to towards the completion of thesis work. Chapter 5 represents the implementation of MBIST controller which is based on modified March C algorithm. Basically the MBIST controller design is based on FSM architecture. Chapter 6 describes the functional verification of MBIST controller's HDL code. It includes the assertion based verification with complete verification planning for MBIST controller design and its implementation. Chapter 7 shows how to setup the simulation environment for design and verification of MBIST controller. Chapter 8 contains the functional verification and simulation results. This chapter also presents the analysis of coverage scores/result for the different coverage metrics.

Chapter 9 contains the conclusion of thesis work.

Chapter 10 describes the future work and recommendation for the further work on this work.

# Chapter 2

# Memory Built in Self-Test (MBIST)

In the present era, the more data dominating applications on System on Chip (SoCs) require large number of embedded memories having different types and sizes to reduce the computation time. The effect of this is increased processing speed. That's why embedded memories are in focus of technology scaling. Due to very small geometries, embedded memories are susceptible to subtle defects [1, 2]. To detect these defects, the MBIST architecture is used for embedded memories testing. The testing of memories is very crucial because of very low yield of memories during fabrication. In case of embedded memories it becomes a tedious task because of no primary inputs and outputs to external of SoC. To test these memories before tape-out of SoC is necessary, this role is done very efficiently by MBIST architecture. A generic MBIST architecture is shown in Figure2.1. The post fabrication testing of such memories is very expensive and time consuming because input and output of such blocks are not easily accessible [3]. The MBIST enables the automatic testing of embedded memory even without coming in touch with outside test environment. The implementation of MBIST architecture depends upon the requirement of testing and the type of memory under test (MUT) for the optimum and efficient testing. BIST technique integrates the functionality of an automatic test system onto the same die as embedded memories. Therefore, test pattern generation and response can be performed automatically.

As we know MBIST enables the high speed and high-bandwidth access to the embedded memory cores [7]. Now days, MBIST is the state-of-the-art technology to test embedded memory. The main drivers behind the development of MBIST are the high rising cost of Automatic Test Equipment (ATE) and increase in the complexity of the designs or products.
Sometimes Memory Built-in Self-Test (MBIST) is also referred as Array Built-in Self-Test (ABIST). This is an amazing piece of logic. Without any direct connection to the outside world, a very complex embedded memory can be tested efficiently, easily and less costly unless carefully designed. In present scenario and in coming next years, embedded memories will be holding most of the silicon area of chip from 60% to 95% [8] in modern SOCs.

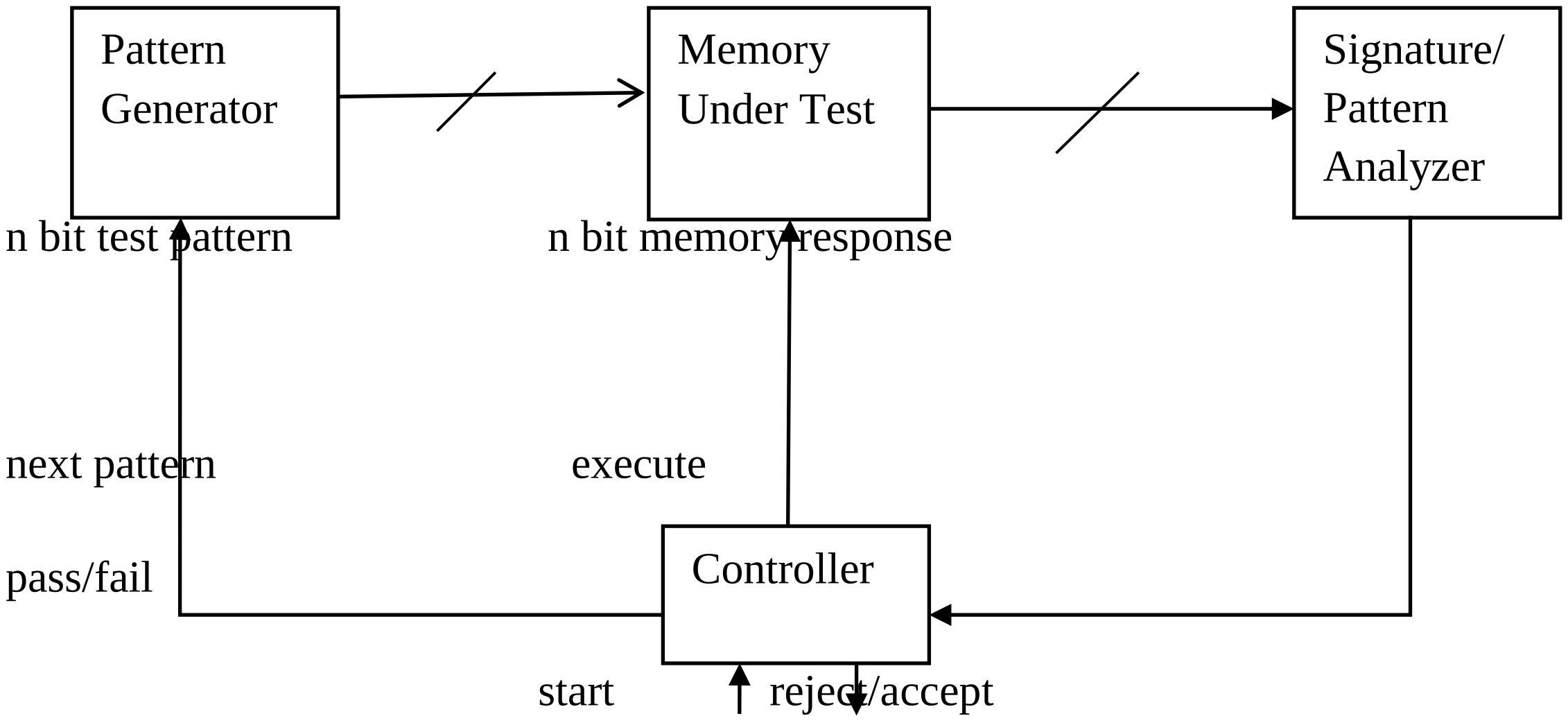

**Figure2.1Generic MBSIT Architecture**

In broad sense a MBIST architecture for its typical functionality requires the following the functional blocks.

**i) Pattern Generator:** To test the memory, it is necessary to have different data patterns which would be read & write from and to memory. Data patterns are generated by pattern generator. The design of pattern generator is based on used pattern type in the selected algorithm to test the memory. There are various algorithms such as March memory testing algorithm, checkerboard, varied pattern algorithms etc.

**ii) Address generator:** Address generator is design to provide the memory addresses to read and write test pattern on those memory addresses.

**iii) Pattern analyzer/comparator:** It analyzes the actual data pattern and the data coming out from the memory after read operation on the memory. Based to the comparison results it generates the pass/fail signals. If both data patterns match then it generates the pass signal which means the data coming out from the memory is exactly same as data written in to the memory. It ensure that memory under test is fault free. But if both data patterns mismatch then it generates the fail signal this indicates that the data coming out from memory is different from the actual data written in to the memory. It shows that the memory under test is faulty.

**iv) MBIST controller:** This is the heart of MBIST architecture which controls the every sequence of signals during the memory testing by generating the proper controlling signals.

There are following advantages to implement the MBIST architecture [3].

a) Lower cost of memory testing because it eliminates the need of external and costly Automatic Test Equipment (ATE).
b) It provides the better fault coverage because the special test patterns are generated specially for the design under test.
c) A MBIST module can be designed to test the multiple memories at a time or for parallel testing which is responsible for shorter test time.
d) Due to chip design of MBIST, it can test the memory at the operating frequency of the system.
e) Testing outside the electrical testing environment.
f) Provides the facility to the consumer to test the circuit or chip before or after mounting in the other application system boards.

With advantages the implementation of MBIST architecture with memory has few disadvantages. There are some disadvantages to use the MBIST architecture.

a) Silicon area overhead
b) Increases the power dissipation in system.
c) It reduces the memory access times because of mux logic for MBIST operation and main operation on memory.
d) Additional pin requirements for the main processor to show the memory testing results.
e) Checking of the correctness of MBIST circuitry is itself necessary because on chip testing can also get fail.

### 2.1 Introduction to MBIST Controller

A MBIST controller generates the controlling signals based on the memory testing algorithm to perform the testing operation by MBIST in correct manner. It controls the generation if data patterns and their generation sequence [9]. For examples, in case of March algorithm it controls the address generator in such a way that address generator generates the address sequence in up direction if the coming pattern is generated for up marching. Otherwise addresses would be generated in down direction if coming pattern is for down marching on the memory's addresses. The controller is responsible to start and stop the testing process of memory. It gets the starting command from top level controller or processor or off chip ATE. After receiving the command from upper/top level processor, it interacts with other components of MBIST circuitry such as pattern generator, address generator, comparator and memory to perform the testing operation in correct manner [10].

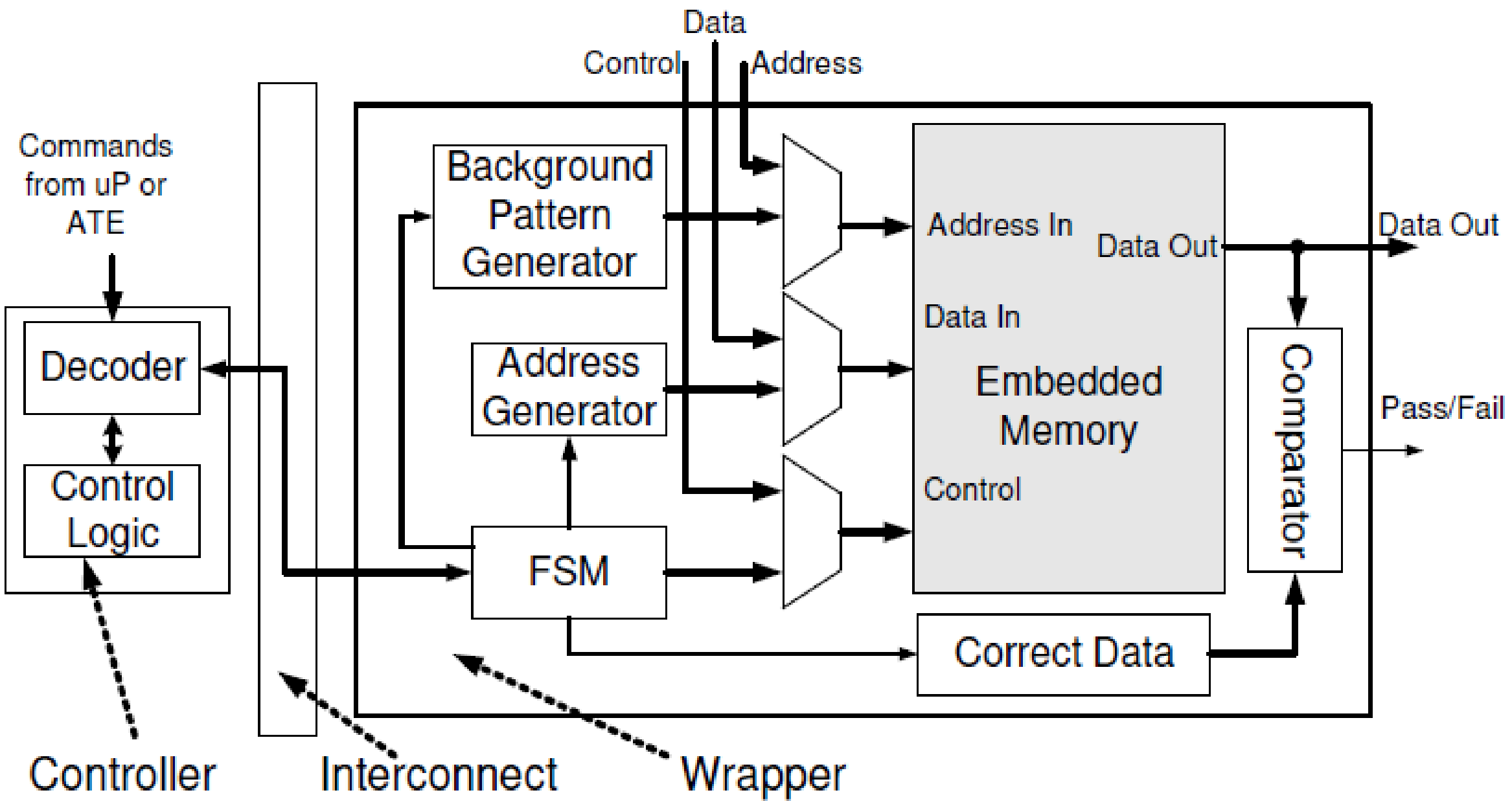

**Figure2.2A MBIST Architecture with controller, other blocks and their interconnections.[11]**

The complete MBIST architecture with detailed architectural components is shown in Figure 2.2. The described partition of the MBIST architecture and terms used in Fig 2.2 are not the standard.

The MBIST controller can be categories in two types mainly.

**i) Programmable MBIST controller:** Such kind of controller design is based on microcode control. The controller can be programed according to different algorithms so it is called the programmable controller. Such controller consist the instruction decoder logic with control logic. The instruction decoder decodes the instruction written as a program in the memory other than the MUT. Such controllers are flexible in nature, they can be modified but they have complex design architecture [3].

**ii) Hardwired or FSM type controller:** Such controllers have hardwired design means the architecture cannot be changed once designed. Controller can be design for a particular algorithm at a time. To modify the controller, complete design has to be restart from the beginning. The Finite State Machine (FSM) implementation of hardwired designs of controller such as shown in Figure 2.3 would be easy and economical. As FSM controller controls the whole MBIST circuitry and defines the correct sequence of events. Different States of FSM represent the different operations in MBIST circuitry. Every state in FSM has its sub-states which represents the all events happening at time when controller would in that state [9]. MBIST architecture based on hardwired or FSM controller has optimum logic overhead but at the same time its lacks in flexibility. Such type of controller is used to detect the known faults in memory [12]. It is the oldest MBIST architecture design technique but still kept developing for the memory testing [3].

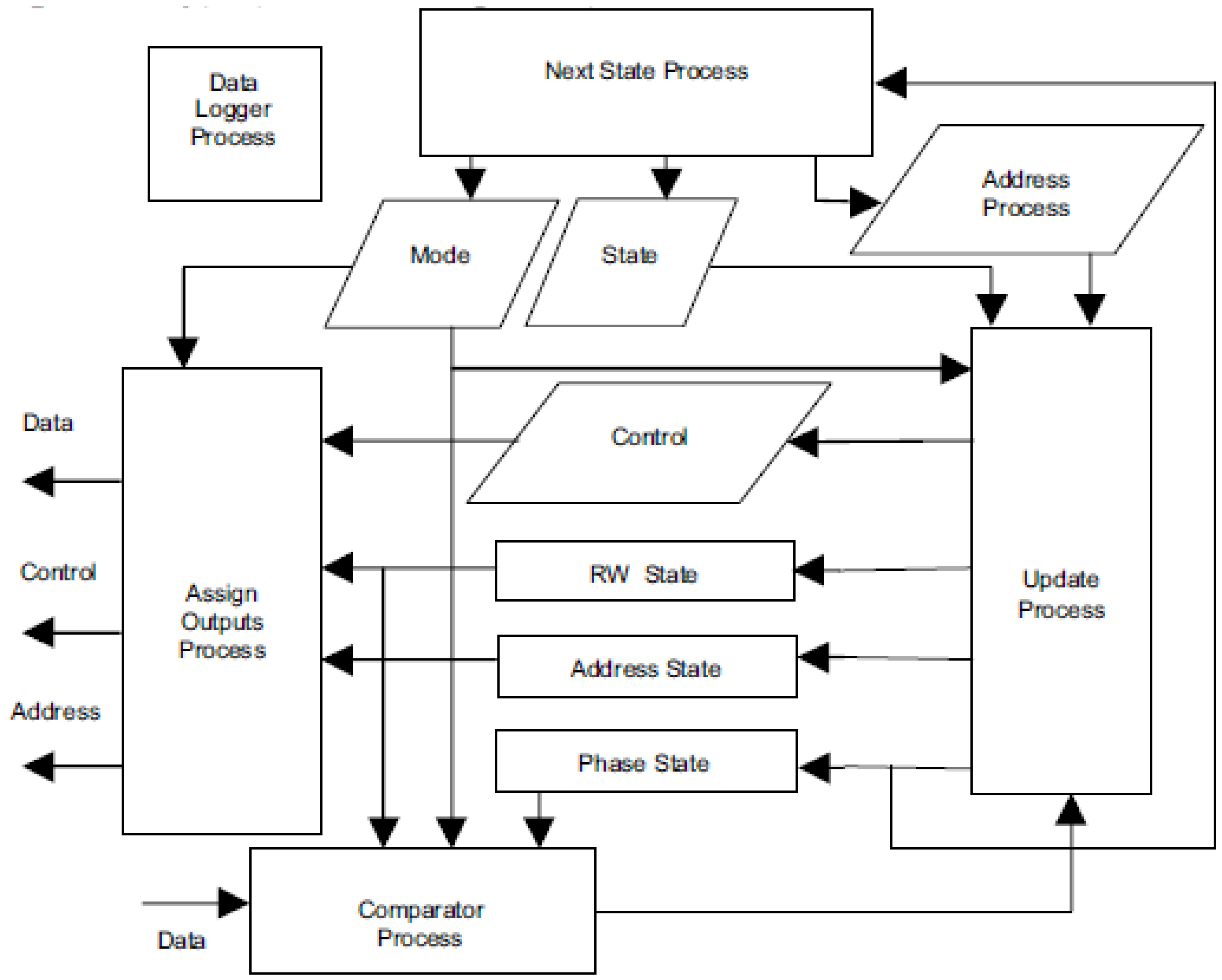

**Figure2.3 A Hardwired MBIST Controller Architecture.[12]**

## 2.2 Different March Algorithms for Memory Testing

To test the embedded memories, designers need the automation means as explained in above section with its advantages. This type of testing requires the predefined approach to test the memories which motivates the developments of different test algorithms for different types of memories. In this process some memories have their testing algorithms which are very much capable to give the better fault coverage compare to other algorithms.

There are several March test algorithm to test the different RAM memories with different fault coverage. It is very much possible that an algorithm may be better to detect particular type of faults but not with all fault detections. March tests are very popular to test the embedded memories because of easy implementation and less testing time. In this section various March tests are explained with their capability to detect the various memory faults.

A complete March test comprises the several March elements. Each March element has sequence of read and writes operations. These operations are applied on each cell of memory in sequence of March elements in March test [13]. In March test, the memory addresses are accessed in two directions in upward direction from $0^{th}$ address to (n-1)th address which called up marching ( ⇑ ) in March testing. Other way is down direction from (n-1)th address to $0^{th}$ address which is called the down marching indicating by ( ⇓ ) in Marching testing. A cell can go under following operations r0, r1, w0, and w1. These symbols indicate the following operations.

a) r0 : Read ‘0’ as data/pattern value from the memory cell.

b) r1 : Read ‘1’ as data/pattern value from the memory cell.

c) w0 : Write ‘0’ as data/pattern value in to the memory cell.

d) w1 : Write ‘1’ as data/pattern value in to the memory cell.

The boundaries of a complete march test of a March algorithm is delimited by the {….} brackets while a March element is restricted by the (…) brackets.

Some March test algorithms [12] are listed below. And their complete description with number of required time cycle and names of faults detected by that particular algorithm are shown in Table 2.

MATS+: The MATS+ algorithm is shown in Figure 2.4 It comprises only three March elements.

{ ⇕ (w0); ⇑ (r0, w1); ⇓ (r1, w0)}
M1 M2 M3

**Figure2.4 MATS+ March test algorithm with March elements.**

M1, M2 and M3 represent the all three march elements.

March B: The March B algorithm is shown in Figure8. It comprises five March elements represented by M1, M2, M3, M4 and M5 in Figure2.5.

{ ⇕ (w0); ⇑ (r0, w1, r1, w0, r0, w1); ⇑ (r1, w0, w1); ⇓ (r1, w0, w1, w0); ⇓ (r0, w1, w0)}

M1 M2 M3 M4 M5

**Figure 2.5 March B test algorithm.**

Some other March algorithms are presented in Table 2.1 [13].

**Table2.1 March Test Algorithms**

| | |
|---|---|
| MATS++ | {⇕(w0); ⇑(r0,w1); ⇓(r1,w0,r0)} |
| March X | {⇕(w0); ⇑(r0,w1); ⇓(r1,w0);⇕(r0)} |
| March C- | {⇕(w0); ⇑(r0,w1);⇑(r1,w0);⇓(r0,w1); ⇓(r1,w0);⇕(r0)} |
| March A | {⇕(w0); ⇑(r0,w1,w0,w1); ⇑(r1,w0,w1); ⇓(r1,w0,w1,w0); ⇓(r0,w1,w0)} |
| March B | {⇕(w0); ⇑(r0,w1,r1,w0,r0,w1); ⇑(r1,w0,w1); ⇓(r1,w0,w1,w0); ⇓(r0,w1,w0)} |
| March SR | {⇓(w0); ⇑(r0,w1,r1,w0); ⇓(r0,r0); ⇑(w1); ⇓(r1,w0,r0,w1); ⇑(r1,r1)} |
| March SS | {⇕(w0); ⇑(r0,r0,w0,r0,w1); ⇑(r1,r1,w1,r1,w0); ⇓(r0,r0,w0,r0,w1); ⇓( r1,r1,w1,r1,w0); ⇕(r0)} |

In the following Table 2.2 some tests are summarized in term of number of required time cycle and types of fault detection.

**Table 2.2 Summarized Some March Test Algorithms [14]**

| Test | Test Time | Covered Faults | | | | | | |
|---|---|---|---|---|---|---|---|---|
| | | **AF** | **SAF** | **TF** | **CFin** | **CFid** | **CFst** | **LCFids** |
| MATS+ | 5n | A | A | - | - | - | - | - |
| March C | 10n | A | A | A | A | A | A | - |
| March B | 17n | A | A | A | A | A | A | A |

In the Table 2.2, n is the number of cells in memory array. A stands for All. AF –Address decoder faults; SAF- Stuck at Faults; TF –Transition Fault; CFin – Inversion Coupling fault; CFid–Idempotent coupling faults; CFst – State coupling faults; LCFids – linked Coupling Faults.

From the above table, it clear that if there is less possibility to has linked coupling faults in memory than March C test algorithm is much efficient in test time and number of covered faults.

# Chapter 3

# Verification

Verification is the process to check whether the intent of the design is kept while implementing it. This chapter describes the various aspects of verification process. In the present era 70-80% design efforts comes under the verification domain. A verification process takes a large amount of time but it can be reduced by automation of various aspects of verification process.

## 3.1 Verification Technologies

Verification process follows two main approaches one is simulation based verification and another is without simulation called formal verification.

**i) Formal Verification:** The approach in formal verification is to check the design's functionality or correctness based on mathematical models. The formal verification tools works on the assertions written to check the functionality where assertions represents the properties of design. Formal verification of complete design is not a feasible task because it becomes much complex and time consuming. So the formal verification techniques are applied on the selected modules of complete design and mainly for the corner cases. These corner cases are not tested during the simulation. In this way formal verification becomes very effective and less time consuming. Formal Model checking and equivalence checking are two main approaches used in formal verification.

**ii) Simulation Based Verification:** Simulation based verification is still the very common and popular verification technology. In a very simple way a test-bench is used to generate and apply the test stimulus to verify the design in simulation. Test-bench is also used to fetch the output from the design corresponding to the applied stimulus. Design which is going to verified is called the design under verification (DUV). While simulating a design it can report the presence of bug but it can never assure that design is bug free. For a normal size design to apply all possible combinations and sequences of stimuli is tedious and time consuming task during simulation. That's the main drawback of simulation verification.

By combining the both approaches the drawbacks of both approaches can be removed. A different verification approach called Assertion Based dynamic verification.

**iii) Assertion Based Dynamic Verification:** It is also called assertion based functional verification (ABV) which includes assertions with simulation and formal techniques to the traditional simulation based functional verification of the RTL designs [12]. ABV firmly depends on assertions to verify functionality and observe functional coverage by putting SVA specifications. Due to this, the verification status is provided by the parallel run of coverage and assertion checking. By using SVA specifications, there is no need to add tool specific semantics during verification or writing assertions.

Assertion based verification is now one of the mainstream for functional verification of designs. The theoretical research to find out new formal methods and industrial practice in hardware and software verification provides the great benefits to the assertion based verification [13]. ABV is also a powerful functional verification approach which increases the efficiency to verify the correct functionality, bug detection and bug removal during the design process. Assertions are the main elements in the ABV which helps to find out the untested areas or corner case in the design and to write the better and directed test cases to test those untested areas. These corner cases may keep bugs which can cause the wrong functional behavior of design. In this way ABV helps to finds out the bugs present in the design. It improves the level of verification from RTL to design specifications [14]. ABV also comprises the coverage property which holds true for a particular functionality if it is exercised by a test.

This technique can be applied to complete design and the assertion helps to write the directed test cased or stimuli for the DUV. In this way it adds the advantages of the above discussed technologies with removal of their drawbacks or disadvantages.

### 3.2 Verification Plan

Mainly verification plan is described as the specification of functional verification and specification of the architecture of test-bench. It also describes the first time success and verification approach of design. The detail functional specification documents for design are written using system specification. It also ensures that the design must meet the time deadline with correct functionality [15].

The role of verification plan is very crucial and important in verification. In traditional verification the complete responsibility is on the hardware designer and his decisions would be simple. He would be verified it up to time permits. It can be correct functionally but for its integration with the system may not be smooth. Later many devices are used to fix the integration problems for the cost of its flexibility [16].

These challenges can be addressed by verification planning. The role of verification planning is to specify the verification, defining first time success, level of verification, tools and strategies, from specifications to features, assertions and coverage points & goals.

#### 3.2.1 Specifying Verification

The specifying verification is all about to decide your verification requirements with proper schedule. It requires complete details of work to be done. This helps to determine the end point or completeness of verification. A document of the design under verification must be there which represents the design specification in well written form before start the verification.

The contents of specification document depend at which abstraction level, the specifications are written. There are two main abstraction levels to write the specification of a design.

i) Architectural level specification.
ii) Implementation level specification down to block/ architecture level.

Later when there are discrepancies between expected results from implementation and results coming out after verification, then it is try to found bugs in implementation or specification used for verification. Specifications must be already prepared before the implementation of design. In other words verification plan is specification documents of design under verification [18].

### 3.2.2 First Time Success

First time success is the success of the all defined properties at least once during verification. It ensures that the functionality defined as properties is verified. Now to get the first time success, it is necessary to identify the conditions under which features should be verified and the expected results for the features defined as properties. Only after successful pass for decided testcases and the satisfaction of coverage metrics, a design can be passed through its verification process.

### 3.2.3 Levels of Verification

The design implementation goes through the various levels like logical partition (synthesis of units, blocks, reusable cores etc.) and physical partition (PCBs, FPGAs, ASICs etc.) so it is most necessary to decide the level of verification.Figure9 shows the abstraction levels of design and their verification requirements/ applications.

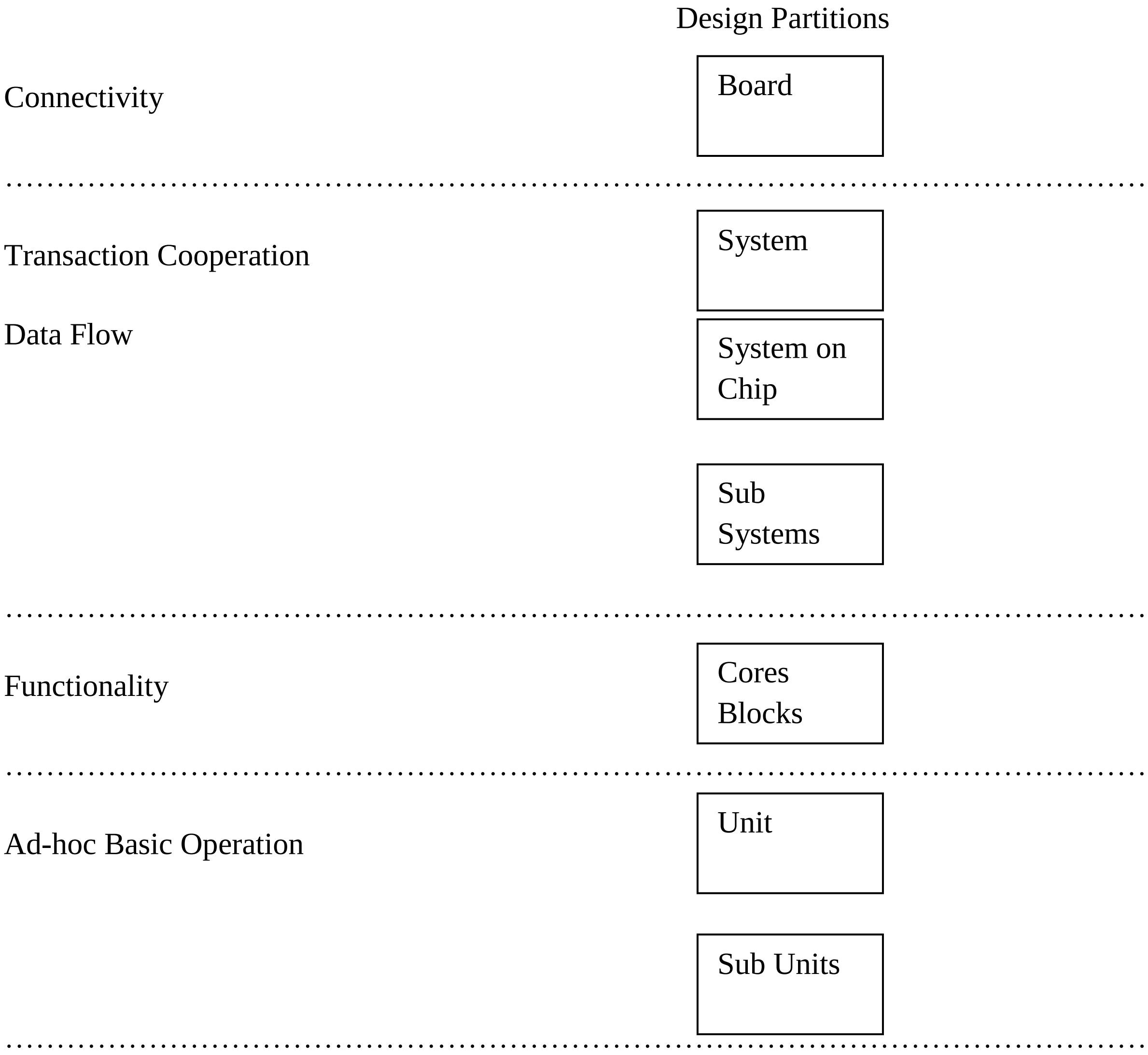

**Figure 3.1 Verification applications at different abstraction level of design**

Before deciding the partition levels to be verified their functionality and interfaces must be stable as much as possible. If interfaces and functionality of the design are changing at those particular levels of abstraction, every times test-benches must be changed accordingly. The partition which is going to be verified, it must have its own specification document.

At unit or sub unit level, the ad-hoc verification operation is required because a specification document cannot be written for this partition of design. The main reason is the interfaces changing with change in implementation process. The designer verifies the design by inserting some assertions and using testcases at this level of verification to check the basic functionality. A large and complex design has many design units but it is necessary to perform the verification at unit level. This increases the controllability and observability of design during verification which helps to provide confidence to validate the design.

The block level verification is done independently because this partition of design has its own independent documentation. A block comprised units of the design. All blocks of design require same effort of verification and it does not depend on the sizes of blocks. The blocks in a design may have different sizes. Design includes the some reusable core blocks which have their own intended functionality in different designs. So they must be verified independently for usage in any design. Here assertions play critical and important role to put restrictions and requirements for such core blocks of design so that the blocks does functionality as intended in that particular design.

At block level verification, the features are verified which are within it. Once verified then it can be assume that verified features will work well at higher level verification. When features verified at block level have the interaction with other blocks then they have to be verified again at higher level to ensure the correct integration of blocks.

When system level verification is done then at that time the functionality of block level design must be perform correctly. At system verification the focus is kept on integration of different blocks or sub systems.

### 3.2.4 Tools and Strategies

In this section it is decided what will be the strategies and tools to verify the design. Selection of tools and strategies depend on the requirement of design module whether it can be verified using formal techniques or simulation techniques etc. A design or its modules can be verified with different verification techniques and strategies. More about this can be found in section 3.1 and chapter 4.

### 3.2.5 Specifications to Features of the Design

This step is the heart of verification plan because complete verification plan revolves around it. It decides what and which type of features will be verified in verification plan. Then it decides what verification strategy should be followed. The specifications from the documentation of design are converted in to the features during the implementation. Now these features are verified against their specifications. Sometime system architects and RTL designer add some extra features to be verified. It is explained [20] how to extracts the features to verify and what are the relevant features at interfaces, functions and then the corner cases.

Most of the interface features can be written to find the answers of the following questions.

What type of transactions can be applied?

a) Transaction's values?
b) What are the protocol violations in design?
c) Interactions between interfaces?
d) Synchronization of transactions?

### 3.2.6 Assertions

The Planning of assertions is very important in a verification plan because different assertions are used to check the more details about the design. When assertions are included in verification process then an assertion language like system Verilog is used to specify the expected behavior of a design in form of assertions. Assertions are checked mainly in two ways one is dynamically and other is statically. When assertions are checked dynamically then it is simply called assertion based dynamic verification. Here, dynamic term is linked to the simulation. There are following main uses of assertions.

i) Clarify specification requirements.

ii) Capture design intent of implementation.

iii) Validate correct operation and usage of design.

There are following two types of assertions.

a) Immediate Assertions
b) Concurrent Assertions

Sometime assertions are also categories based on whether these are implemented by design engineer or verification engineer [21].

While implementing the assertions it is necessary to define the place where to embed these assertions. Assertion can be embedded in different places like in test-bench, in the source code or by using bind statement as a separate file. Besides that, the following information should also be specified.

a) Functionality which is going to be verified by assertion.
b) Where the assertion is placed whether it is external or internal to design.
c) Type of assertion whether it is immediate or concurrent assertion.
d) Whether assertion is implemented by design engineer or verification engineer.

### 3.2.7 Coverage Collection, points and Goal

To obtain the coverage results, first types of coverage are defined then cover points in the design and the goals for the various types of coverage metrics. The coverage points are defined for the functional coverage metric. So the cover points must be listed after the planning. The list of cover points should consist following points.

a) Name of signal or expression to cover

b) Motive to cover

c) Placement of coverage groups whether it is internal or external to design.

d) Defined the coverage goal for various coverage metrics/ groups.

### 3.3 Coverage

A verification engineer can determines the level or depth of verification by coverage analysis of the design under verification (DUV). It can be analyzed that which part of HDL code of design is tested or not during simulation by monitoring the execution of code. The coverage analysis highlights the uncovered code portion[22] thus It provides the clear understanding where to put the effort to test untested design functionality to achieve 100% coverage which is desirable for any design. Achieving 100% coverage cannot give 100% surety that design is error free. It provides the systematic approach to attain completeness of verification.

For this purpose, code coverage and functional coverage metrics are used to verify the design in HDL. The code coverage comprises the several coverage metrics and can vary based on used tool for coverage analysis. The basic idea behind these coverage metrics is to cover up the design structure completely written in HDL. While the functional coverage focuses on the functionality of the design. Till now there is not such a metric which is accepted for complete and reliable verification coverage.

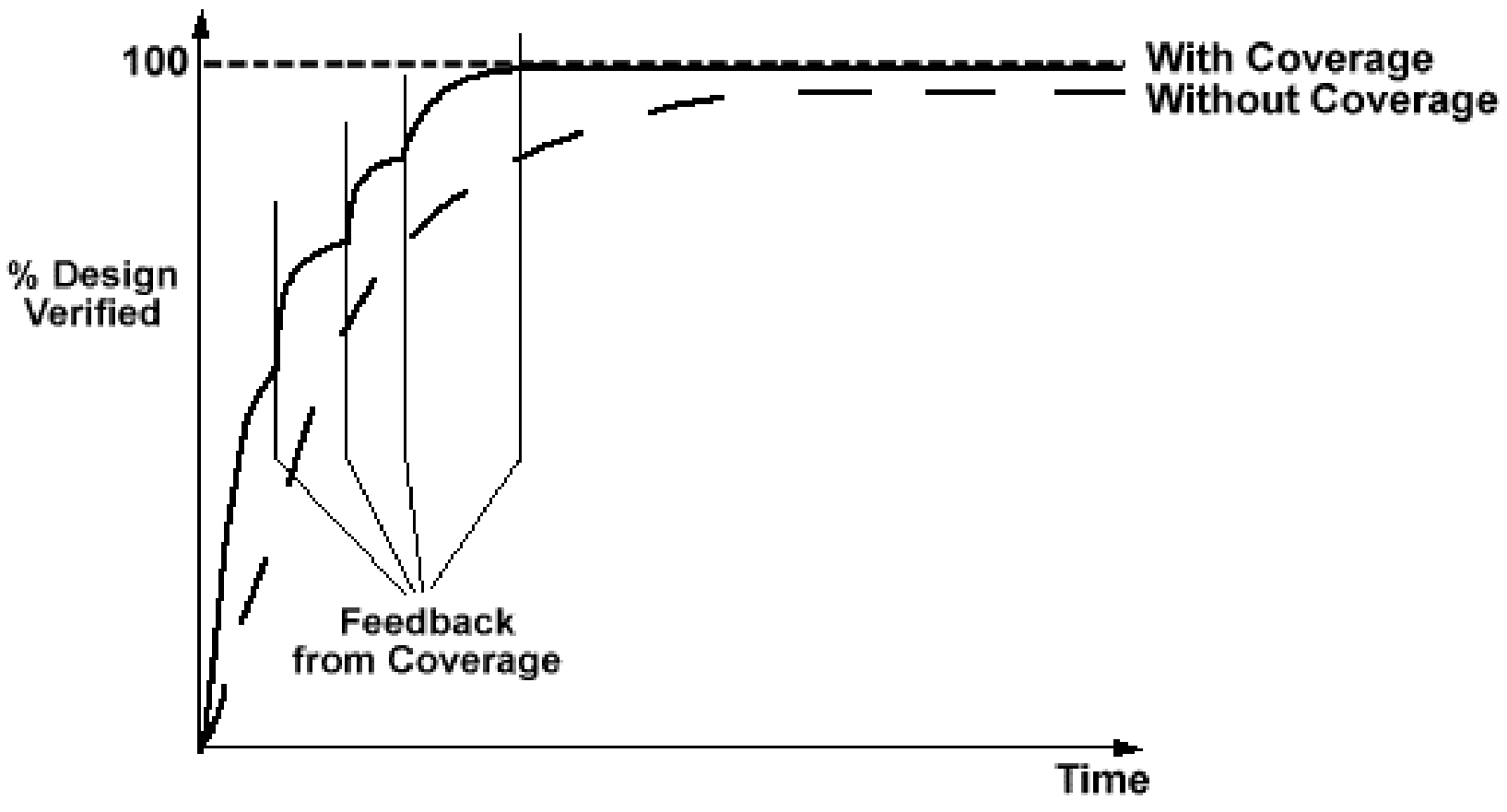

**Figure3.2 Effect of applying coverage analysis on design time and coverage .[22]**

The improvement in the verification in less time is possible by using coverage analysis is shown in Figure 3.2.

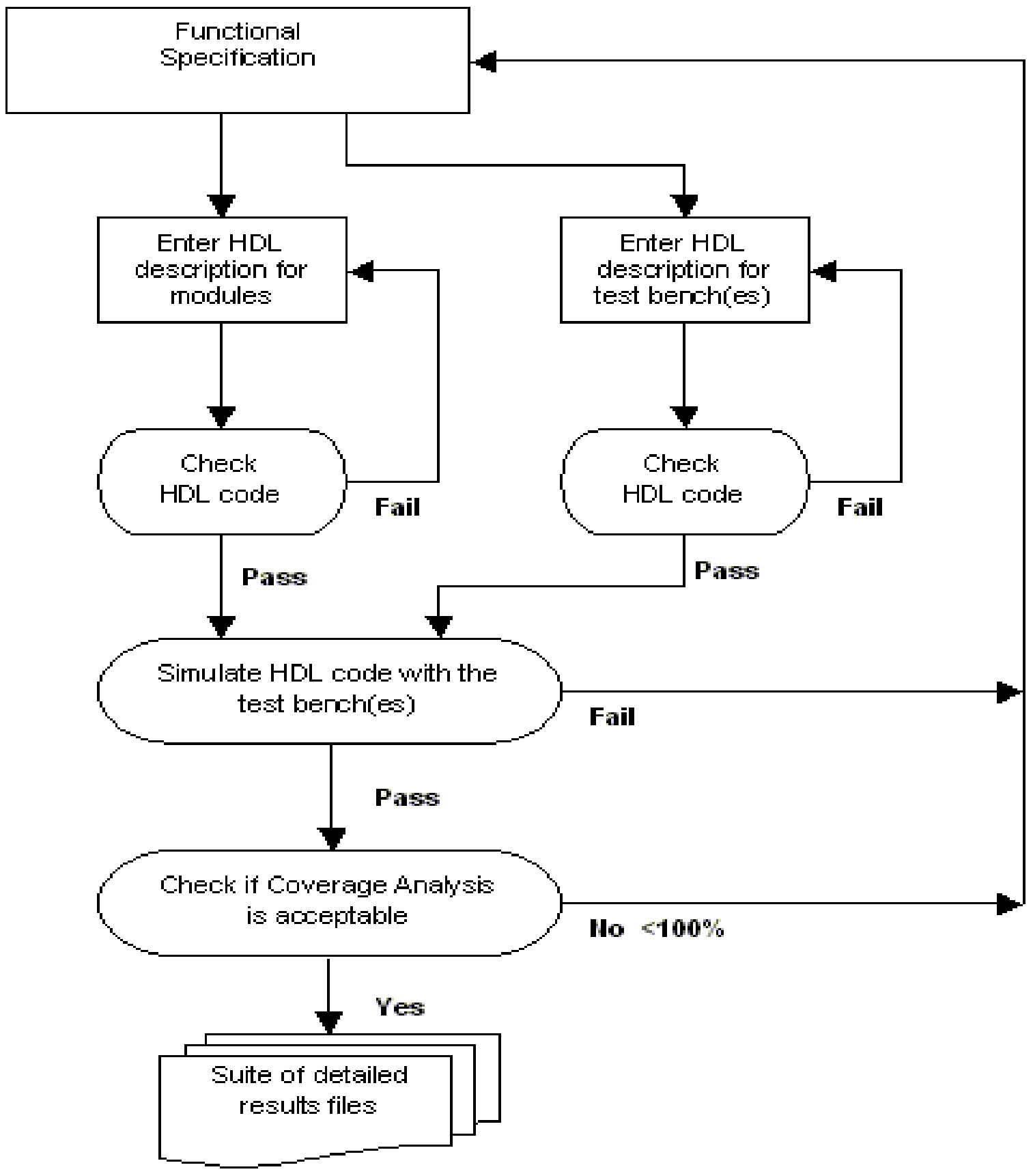

**Figure 3.3 Coverage analysis is incorporated in the design flow. [22]**

Figure 3.3 shows how and at what stage of design flow coverage analysis task can be incorporated. It shows the conditions to pass the coverage analysis, these conditions or constraints depends on designer or verification engineer. These people set the acceptance levels of coverage.

A coverage analysis task can be complete with a coverage analysis tool which shows the coverage score for a particular test-bench. Based on high coverage score, test-benches can be ranked well than others. The test-bench which produce the highest coverage score with in less simulation time is assumed the best suited test-bench for design. In this way a designer can optimize the test suits.

### 3.3.1 Code Coverage

Code coverage has several types of coverage metrics. Sometime different names are used for similar types of coverage metrics [23] depending upon the understanding and used coverage tool. Most of them are explained below.

**i) Line Coverage:** Line is one of the simplest structures of HDL code. A line of system Verilog code is said to be covered if it has had transaction on it. An event may or may not occur during transaction at that line but line count as covered. [24]

**ii) Statement Coverage:** This coverage metric only counts the executable statements from HDL implementation of design during coverage analysis. A line may contain more than one statement.

**iii) Block Coverage:** Block coverage is also called segment coverage of HDL code. Its measuring unit of code is a sequence of non-branching codes which is executed at the same simulation time [23]. Block coverage reduces the recording units for coverage analysis. Mostly system tool records as block coverage and display results in term of statement coverage but there is a slightly difference between them.

**iv) Branch Coverage:** Branch coverage metric is also called decision coverage matric. This metric involved the control flow through the HDL code during simulation and can be represented as control flow graph (CFG) to code [26].

**v) Path Coverage:** Path coverage measures the coverage of all paths present in the HDL code. A path is defined as a unique sequence of branches or decisions from the starting of a code section defined in HDL to the end of it [26]. A path in HDL code must contain an edge which is not included in other paths. Path coverage score is based on to cover multiple sequential decisions.

**vi) Conditional coverage:** Conditional coverage or multiple condition coverage [27] is sometimes called expression coverage because the conditions are evaluating based on variable or expression in the conditional statements. It provides the coverage statistic for variables and expressions. It is a very important and critical coverage metrics because it can find the errors in the conditional statements that cannot be easily found by any other coverage analysis [23].

**vii) Event Coverage:** Most HDL simulators are event-driven. Therefore, it is necessary to care about the possible events in a design. Events are associated with the change of a signal. This coverage metrics is very useful when there are too much control events in the design.

**viii) FSM Coverage:** In code coverage point of view, the FSM coverage metric cover number of traversed states in FSM design during the simulation. Here FSM coverage is defined as language-based code coverage for the HDL code just to show whether the all design states are traversed or not. It is most important coverage metric for the FSM based design because it found out most of the design bug due to its closeness to the behavior of design space.

### 3.3.2 Functional Coverage

Metrics defined in this category such as toggle coverage, Sequence and transition coverage of FSM and assertion coverage are related to computation performance of HDL code rather than its structure [26]. The main motive to define such metrics was to exercise each functional scenario of the design after HDL implementation. In FSM designs while performing functional coverage, the coverage monitor looks for error in state transitions and event sequencing mainly. Assertions which interpret design hardware functionality in system Verilog language are considered the special case to monitor in functional coverage analysis.

There are following metrics which are considered as functional metrics in this work.

**i) Toggle Coverage:** Sometime toggle coverage is also called as variable coverage [23]. It measures that each bit in the nets and registers or bits of logic change their polarity during simulation and not stuck at one level [22]. Toggle coverage metrics is considered as first functional coverage metric because without tested a bit properly function coverage target cannot be complete.

**ii) FSM's Transition and Sequence Coverage:** As functional coverage point of view of FSM design, the transition coverage metric and sequences coverage metric are considered as functional coverage. The transitions and possible sequences in HDL code among the states of FSM depend on the functionality of design.

**iii) Assertion Coverage:** Assertions are embedded within the HDL code of design annotates the functionality of design and their main purpose was to generate the assertion coverage metric. This assertion coverage metric covered the successful execution of assertions written form specification documentation of design [6]. Assertion coverage is generated by defining cover groups or cover properties. Cover groups include more than one cover point usually where expected value of data signals are grouped or written in ranges called bins. Every cover group comprises its bin and coverage analysis shows covered and not covered bins. Cover properties also another functional coverage technique. A cover property can be defined for the property defined as an assertion that's why cover property is more like assertion. The analysis result for cover properties shows the number of hits, matches and no matches.

### 3.3.3 Completeness of Verification

Coverage analysis is the only way to know that how much a design has been verified and completeness of verification.Figure3.4shows how to proceed to get the desired code or functional coverage. The goal of verification is always high functional coverage and high code coverage as much as possible.

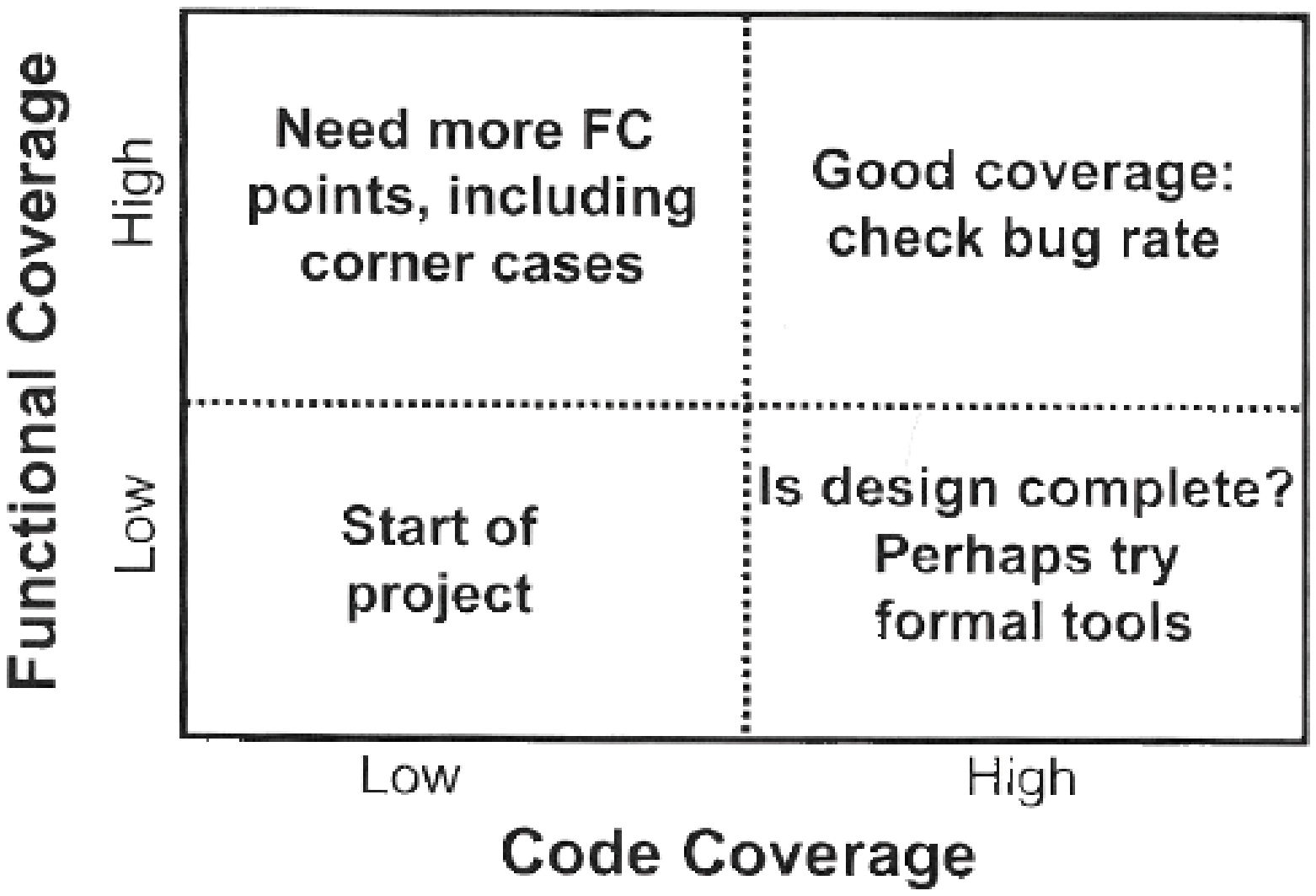

**Figure 3.4 Functional and code coverage analysis basis on coverage score.[28]**

# Chapter 4

# Tools and Languages

The presented thesis work is the design and verification of MBIST controller. The Synopsys-VCS® is used as the verification tool and System Verilog is used as the Hardware Design and Verification Language (HDVL).

Initially, there were two main verification languages ‘e’ and ‘Vera’. These languages introduced the constraint randomization and object orientation features in to the traditional verification. But disadvantage with any Hardware Design Language (HDL) and using ‘e’ or ‘Vera’ as verification languages is that these are completely different languages and both must be known. But a lot of work is already done in ‘e’ and ‘Vera’ [29].

Now System Verilog which is main stream language for both hardware design and verification. System Verilog is a verification language with all important and crucial features of HDLs. That’s why it is called the HDVL. The System Verilog as a HDVL is described in next section.

## 4.1 System Verilog

It is well known that System Verilog language is extension of Verilog HDL. But in real scenario it is the merger of well know programming languages (C, C++), HDL (verilog) and verification languages (e, Vera & PSL). System Verilog comprises additional features like enumerated type, structs types, typedefs, type casting and various operators, operators overloading, control flow statements . It also includes the features of object oriented programming constructs like classes, dynamic object creations etc. Features found in concurrent programming languages (Java) are also added to System Verilog.

System Verilog language is design by integrating the best of all these languages mentioned above into one language. Now design and verification engineers are able to work with single language. Due to use of same language throughout design flow, the execution of design, assertions, test and coverage can be done using single kernel by EDA tools. Because the use of same language to write design code, assertions and test-bench, no special APIs are required to accessing each other [30]. The most important and crucial features added to system Verilog except HDL features are needed for verification.

There are following reasons behind system Verilog to become the first and main choice of design engineers.

a) Reduction in number of coding errors due to reduction in the design code.
b) A complex functionality can be represented in concise and easier way to read and reuse as RTL code representation.
c) Very close representation of HDL code to actual hardware.
d) Mismatch reduction between functionality of RTL representation and gate level functionality after synthesis.
e) Use of single language throughout design flow. (RTL models, test programs, bus functional model, reference model)

System Verilog extension to the Verilog for verification specifically includes the following features.

a) Constraint and biased random variable generation
b) Layered test bench generation
c) User defined coverage points or groups
d) Assertions Based Verification
e) Functional coverage
f) Object Orientated features
g) Application Programming Interfaces (APIs)

In 2005, System Verilog became the IEEE Standard as a HDVL language. The APIs of system Verilog used to interface the other language models (like System C) with the System Verilog models. It also helps to process and extracts functional & assertion coverage information.

There is a verification methodology proposed by Synopsys is called Verification Methodology Manual (VMM) and it completely supported by system Verilog [31]. This thesis work used system Verilog language as for both design and assertion based dynamic verification.

### 4.1.1 System Verilog Enhancements to Verilog

System Verilog enhances the Verilog's capability to represent the hardware designs as Register Transfer Logic in several ways.

**i) Enhancements in Data Types:**

a) *typedef :* System Verilog provides the facility to the users to define their own data types. There are some examples listed below which shows the enhancement in data types.

**typedefintergersigned**sint_d;

sint_d a; //a is sint_d data type.

**typedefenum logic**{s0,s1} state;

state wait;

**typedefstruct** {

logic [1:0] signal0;

logic [31:0] data; } word_data;

**ii) Variable Enhancement:**

a) *Logic:* It is variable type in System-Verilog which is similar to reg but it is four valued variable such 0, 1, x, and z whereas, reg is two valued variable 0 and 1.

**iii) For Loop Enhancement:**

System Verilog also allows the multiple for-loop control variables which are local variables to the for-loop. Ex.

**for (int a=0, b=a+const; a<=10; a++, b++)**

**iv) Task and Function Enhancements:**

System Verilog improves the task and function features in several ways from the Verilog HDL.

There are following enhancements in tasks and functions as compare to task and function in Verilog.

a) Function declaration as **void**.
b) **return** can be used to return a function value.
c) Function arguments can be any data type, array, structures, unions and user-defined type.
d) Formal arguments for input and output.

**v) Module and Interface Instances:**

In Verilog, the ports connection and instantiation of modules are verbose whereas System Verilog provides the short cuts of these. There are two types of short cuts one is dot-name and other is dot-star. In dot-star all ports having same names are connected to each other.

**Module A1 (.*, .clk(mclk), .ad () );** //ad is not used.

The ports having different names are connected using dot-name approach.

There are many more enhancements to Verilog can be found in system Verilog documentations. [31]

### 4.2 Synopsys VCS

VCS is used as a high performance compilation, simulation and most important as a verification tool in parallel to simulation of design. VCS provides all these features on single open native platform. VCS delivers the fast and high capacity simulation for RTL functional verification which fasten the overall system verification. VCS's simulating and debugging features assures the validation of the design. VCS provides the complete package of verification which supports the languages other than Verilog and System Verilog. But System Verilog and VCS both are very much compatible to each other. VCS in its VCS-MX[31, 32] mode also supports mixed language designs like with VHDL.

VCS is supported by the multicore facility due to this it reduced the verification time comparatively other verification tools. It runs the design, test-bench, coverage, assertions and bug finding process in parallel due to multicore technology.

The some main components [31] of VCS are following.

**i) System Verilog:** VCS supports the System Verilog 3.1a except some features of System Verilog. Mainly system Verilog adds the new design, assertion and test-bench constructs.

**ii) System Verilog Assertions (SVAs):**VCS supports and very much compatible with System Verilog Assertion like with Open Vera Assertions.

System Verilog Assertion performs following tasks.

a) These can test the Verilog, VHDL, System Verilog and mixed HDL design codes using VCS and VCS-MX.
b) The results produced by the assertions can be viewed with DVE.
c) These can be monitored and controlled as part of design code and System Verilog test-bench.

SVAs have two following directives.

a) Assert -This *assert* directive is written to define a property which represents a functional characteristic of the system. Such properties are specified in form of temporal expressions most of the times. These temporal expressions represent the complex timing and functional behavior of the system.

b) Cover -A *cover* directive is used to verify that a property or sequence which is defined in assertions was covered or not. Based on whether a coverage expression matched or failed to match, the cover directive generates the report of number of times an asserted property got success or failed during simulation. More than one matches or failures are possible in single attempt. In case of multiple matches, a counter is incremented at every match.

**iii) Discovery Verification Environment (DVE):**DVE is the graphically debugging environment of VCS. In DVE, one can trace the signals of only which are needed to check. After looking into wave forms one can find the bug by comparing the waves and debug. The test-benches can be formed based on the waveforms outputs. The default DVE window is shown in Figure13. This default format of DVE Top-Level window can be changed.

**iv) Coverage Metrics built in VCS:** The built in coverage analysis functionality of VCS comprises nearly all aspects of coverage analysis. This analysis includes condition, toggle, line, Finite State Machine (FSM), path, branch and assertion coverage. This every coverage is represented in the form of metric and in these metrics coverage score determines the quality of design during verification. If the coverage is not satisfactory then focus will be on creating some new or additional test cases. Just one time compilation is enough to run the simulation and coverage analysis and to generate results.

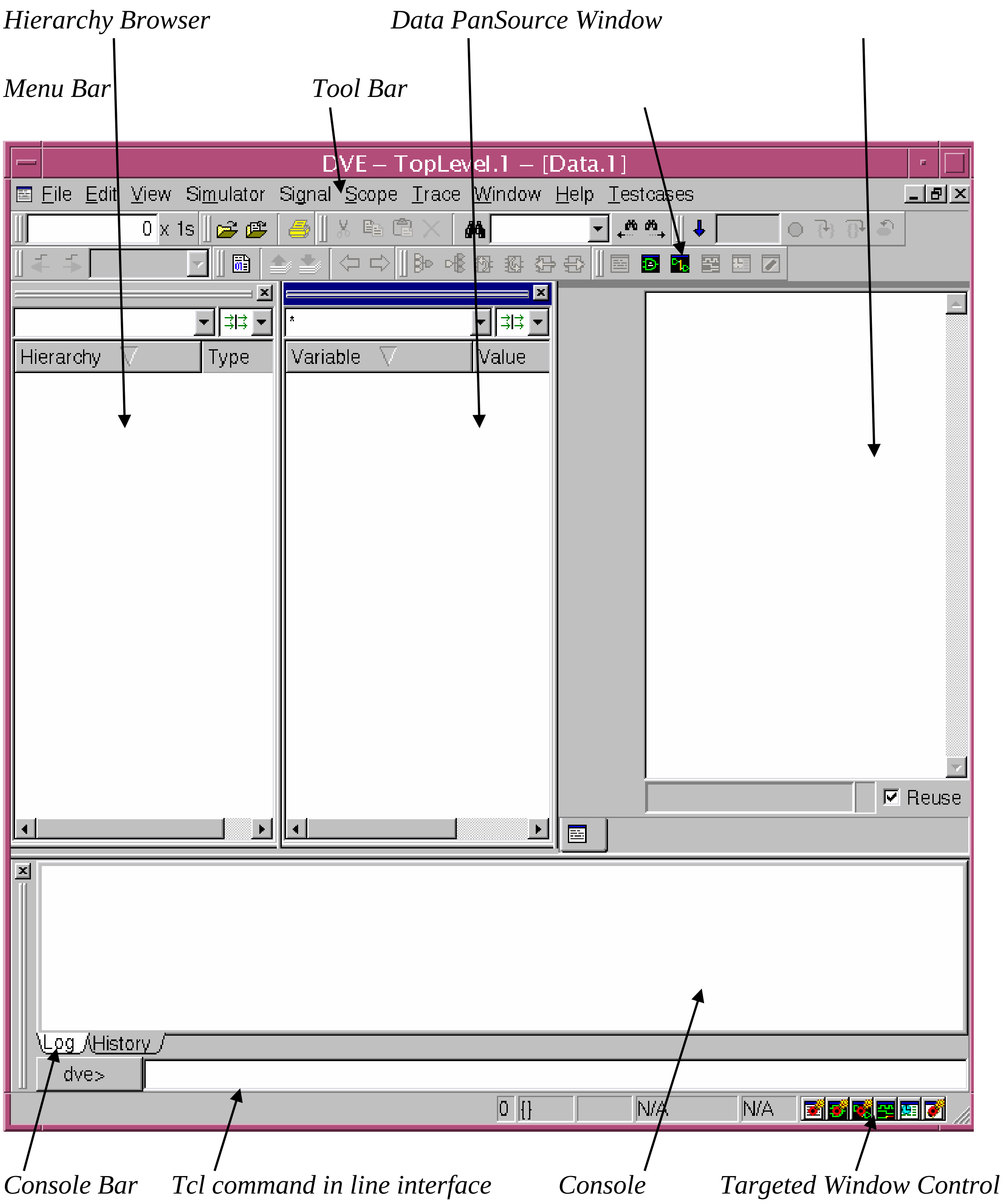

**Figure 4.1 Synopsys VCS® GUI Window frame.**

**v) Direct C/C++ Interface :** If C/C++ like functions are used with in Verilog HDL design code , then such required interfaces are directly provided by system Verilog. There is no need to write separate Programming Language Interfaces (PLI) to interface such functions or modules. VCS has the capability to recognize such functions automatically and to compile & simulate them.

**vi)Incremental Compilation:** In VCS once a design has been compiled then in the next compilation VCS compiles only the part of design code which is modified only. For the rest design it gets the data from the previous compilation data. This is the incremental compilation. The csrc subdirectory stores the compilation time files and results.

**vii) Mixed Signal Compilation:** It is very challenging to verify digital and analog modules together because it required such verification architecture which supports both digital and analog designs simultaneously. Here VCS AMS test-bench provides the solutions to this problem. The traditional digital verification techniques are used with some modification so that it can drive analog IPs like ADC/DAC, clock generator etc.

The more details about mixed signal simulation with VCS can be found in Discovery AMS documentation.

# Chapter 5

# Implementation of MBIST Controller

The controller generates the control signals for other components of MBIST circuitry such as patt_g for data/pattern generator, en to start the address counter inside the address generator, rw for Read/ writes generator and en for signature analyzer and other components too. The top level block diagram of MBIST is shown in Figure 5.1. Mainly the controller handles the test sequence and its result corresponding to memory output. After assertion of T_mode signal from higher level processor or controller, MBIST controller generates the control signals corresponding to the pattern generator, address generator, read/write and signature analyzer. Controller and pattern generator control the up or down address sequence by generating the control signals for address generator. The generation of March pattern is based on the '0' or '1' value of control signal patt_g whether it is marching 0 or 1. Controller generates the control signal rw to the read/write controller for the reading or writing operations from and into the memory [9, 10].

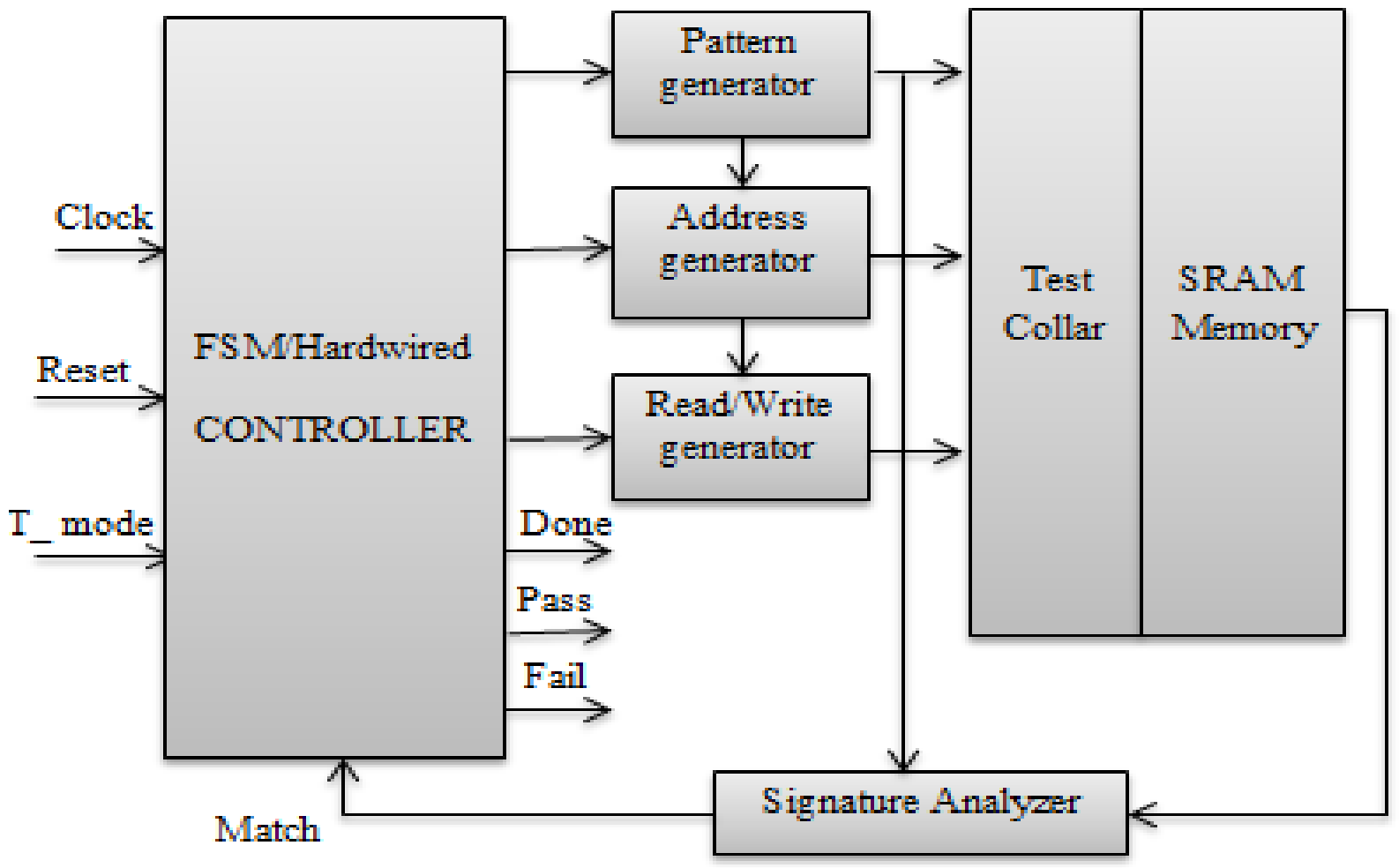

**Figure5.1 MBIST architecture with FSM controller.**

Match signal is the output of signature analyzer and input to the controller. Based on the value of match, controller asserts the pass or fails if match is '1' or '0' correspondingly. Only after completion of March test sequence, controller asserts the done signal. The controller is designed based on March C algorithm [10, 14] which is capable to detect SAFs, AFs, TFs and CFs (except linked CFs) with an additional pause element to detect the data retention faults (DRFs) in memory.

**5.1 Implementation of March Algorithm**

The March C algorithm consists six march elements say M1, M2, M3, M4, M5 and M6. The pause element M7 is added to test the retention time of SRAM under test. It does not access the memory while it pauses the marching of element to check whether memory is able to retention the written data or not for a particular time.  Pause element is added just after the write element to check which memory cell is not capable to retain the same written data after a particular time. During pause the controller does not allows the execution of the original march test sequence of read/write operations on memory under test.

$$\{ \underset{M1}{\Updownarrow (w0)}; \underset{M2}{\Uparrow (r0, w1)}; \underset{M3}{\Uparrow (r1, w0)}; \underset{M4}{\Uparrow (r0, w1)}; \underset{M5}{\Downarrow (r1, w0)}; \underset{M6}{\Updownarrow (r0)}; \underset{M7}{\Updownarrow (pause)} \}$$

**Figure 5.2 March C algorithm with pause element.**

Here five read March elements are encountered and each encounter is capable of detecting faults in memory. The controller's implementation is based on March algorithm shown in Figure 5.2 which is capable to detect SAFs, AFs, TFs, CFs (except linked CFs) and data retention faults (DRFs). The March C algorithm includes 6 march elements say M1, M2, M3, M4, M5 and M6. But here in Figure5.2, the implemented algorithm comprises an additional element pause say M7 to check the memory data retention time through this detection and identification of data retention faults are possible.

**5.1.1 Various Fault Detection by Read Operations in March Elements**

By assigning F1, F2, F3, F4, F5 and F6 syndromes [10] to the all five read and pause element operation  correspondingly to distinguish whether an operation detects a particular type of fault or not. The 1 value of any syndrome shows that the corresponding operation is able to detect that particular fault and 0 value shows that fault could not be detected by this operation. The value of syndromes with different memory faults is shown in Table 5.1.

It is clear that all read operations are able to detect the address faults because all syndrome value is 1 for AF. The March element in March C algorithm having a read operation also consists an opposite write element except M6. Due to the presence of defect, [11] more than one address is selected and the different logic values are read at the same time on the bit-line. And the bit-line gets the undefined logic value(x). All syndromes value are also 1 for coupling faults except F6, it means all five read operations are able to detect coupling faults. But all syndromes values are 1 for data retention faults (DRFs) it means all read and pause operation can detect the retention faults in memory but the only 1 value of F6 syndrome can assure the detection of DRFs.

**Table 5.1 Fault detection Syndromes corresponding to all five read and pause element in March algorithm.**

| Faults | Syndromes | | | | | |
|---|---|---|---|---|---|---|
| | F1 | F2 | F3 | F4 | F5 | F6 |
| SAF(0) | 0 | 1 | 0 | 1 | 0 | 0 |
| SAF(1) | 1 | 0 | 1 | 0 | 1 | 0 |
| TF <↑ 0> | 0 | 1 | 0 | 1 | 0 | 0 |
| TF <↓ 1> | 1 | 0 | 1 | 0 | 1 | 0 |
| AF | 1 | 1 | 1 | 1 | 1 | 0 |
| CF | 1 | 1 | 1 | 1 | 1 | 0 |
| DRF | 1 | 1 | 1 | 1 | 1 | 1 |

### 5.2 FSM Implementation of MBIST Controller

The Figure 5.3 shows the implemented FSM structure of MBIST controller which is easy to implement as a FSM and very feasible if no further improvement is required in MBIST controller design. The above FSM state diagram consists of several states and every state represents a particular march element except state idle. Initially controller rests at Idle state and keeps other the MBIST circuitry in idle until t_mode signals is asserted. But after getting value '1' at t_mode signal, controller starts traversing the next states to generate required pattern for read or write operations from and into the memory cells. Now in offline testing MBIST controller takes the controls of memory from higher level processor.

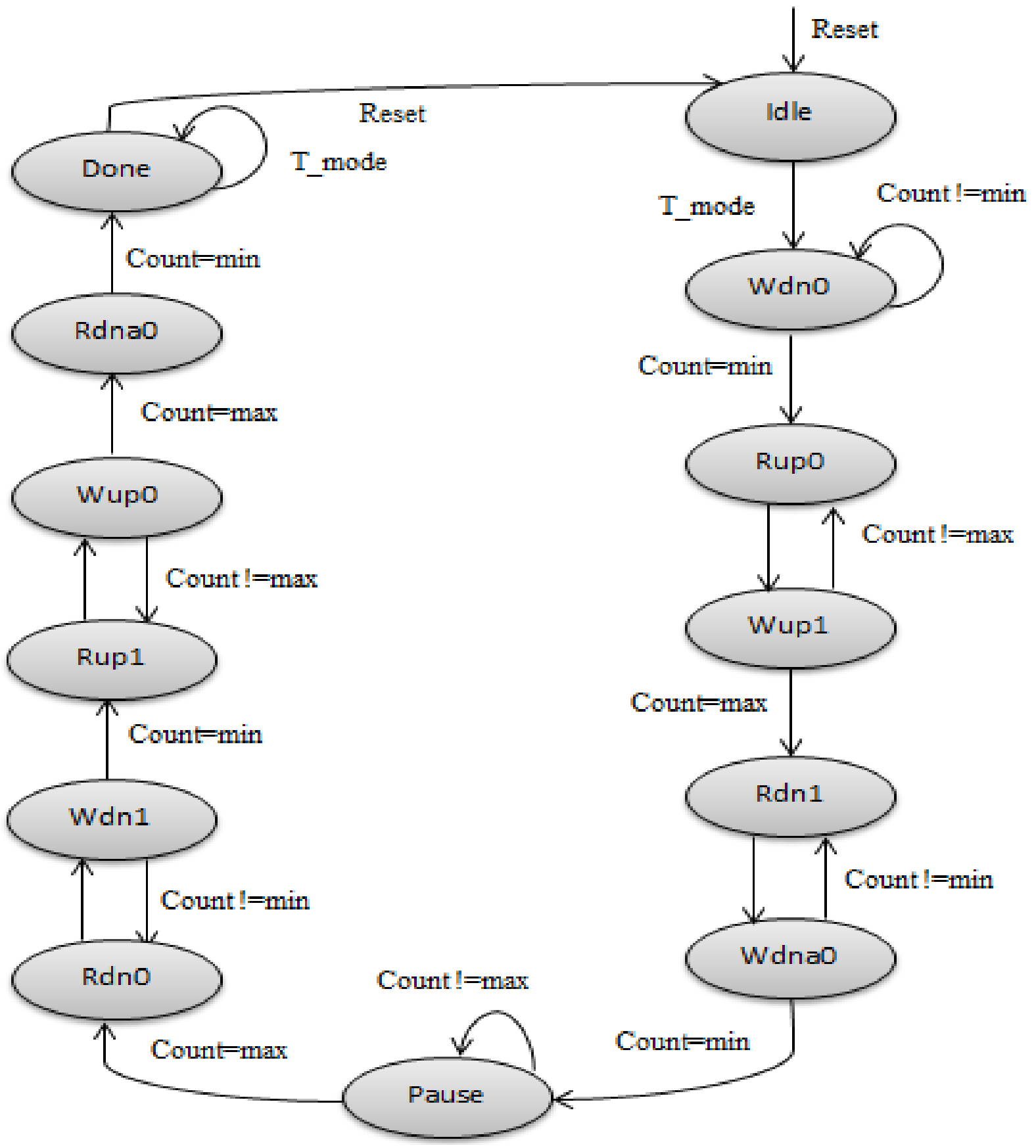

**Figure5.3 Finite State Machine Diagram of MBIST Controller.**

At every state a read/write signal 'rw' gets a value '1' or '0'. State Wdn0 comes when t_mode signal is asserted high and at every maximum and minimum count value the present state changes to the next state as shown in Figure 5.3. The maximum value of count depends on the address width of SRAM memory. In normal condition if a state transition happens, it means a March element gets complete. At first March state Wdn0, 'rw' signal is kept at '0' and read/write generator write data in memory until count value gets minimum. Here 'W' represents the write operation, ‘dn’ shows the down marching and ‘0’ shows that marching pattern is 0. During up marching**,** count starts from minimum address value to the maximum address value and state changes only when count gets its maximum value unless t_mode and reset signal change. In case of down marching, state transition occurs only when count gets its minimum value. Pause state is inserted between a write state (Wdn0) and a read state (Rdn0). During pause, controller just holds the marching operation on memory for the specified time. The March testing follows the read-test-write sequence that’s why every read state in FSM design comprises the test result as a signal pass /fail. After completion of test controller enters into the Done state and controller asserts the done signal to show that the test is completed. The FSM shown in Figure5.3 is implemented using System Verilog in Synopsys VCS tool.

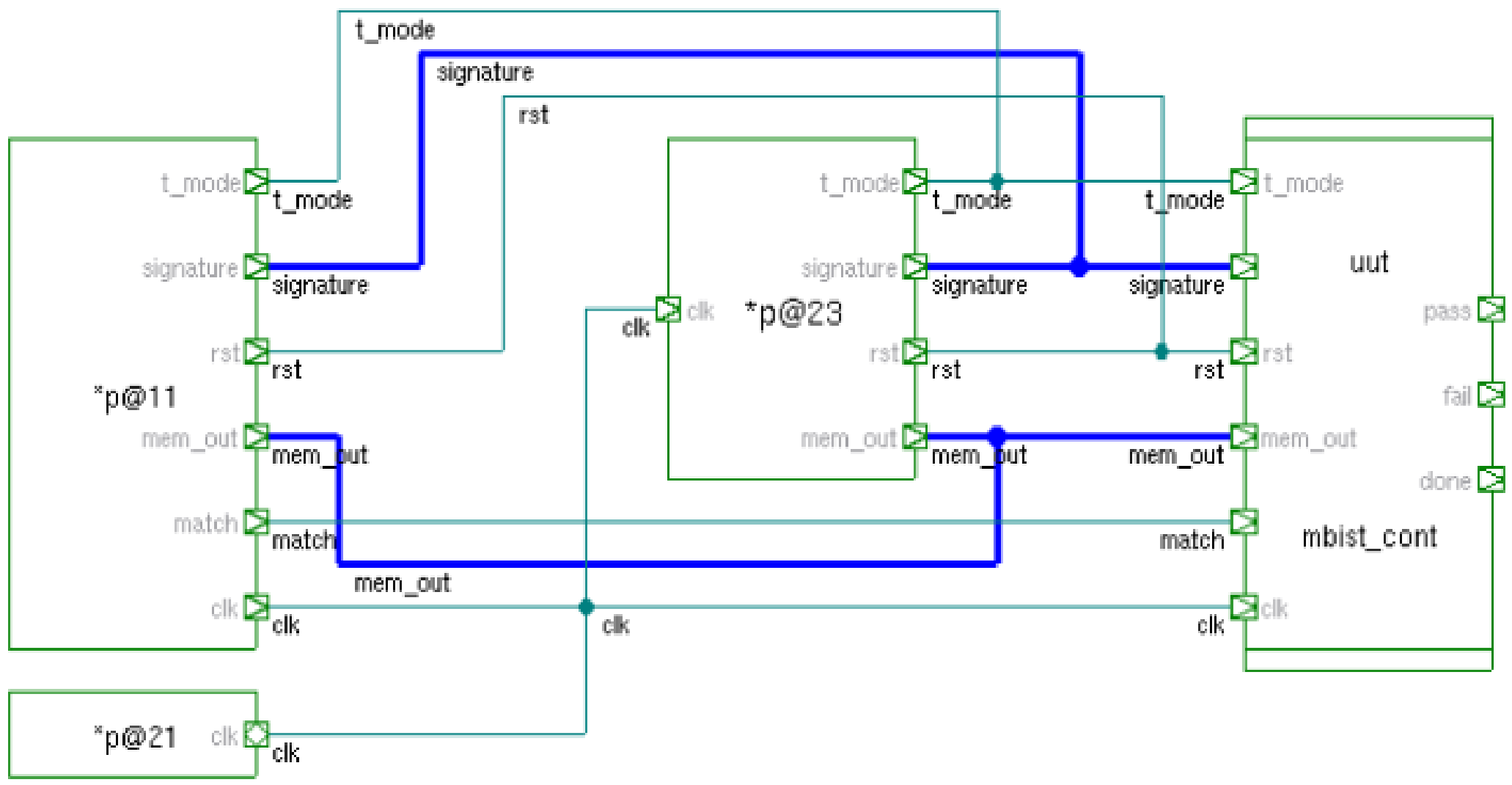

**Figure 5.4 MBIST controller schematic.**

The Figure 5.4 shows the RTL schematic of MBIST controller design generated by Synopsys VCS®.

Figure 5.5 shows the simulation waveform where the functionality of designed MBIST controller is tested for a memory fault inserted which is detected by the 0 marching pattern in rdna0 state of controller.

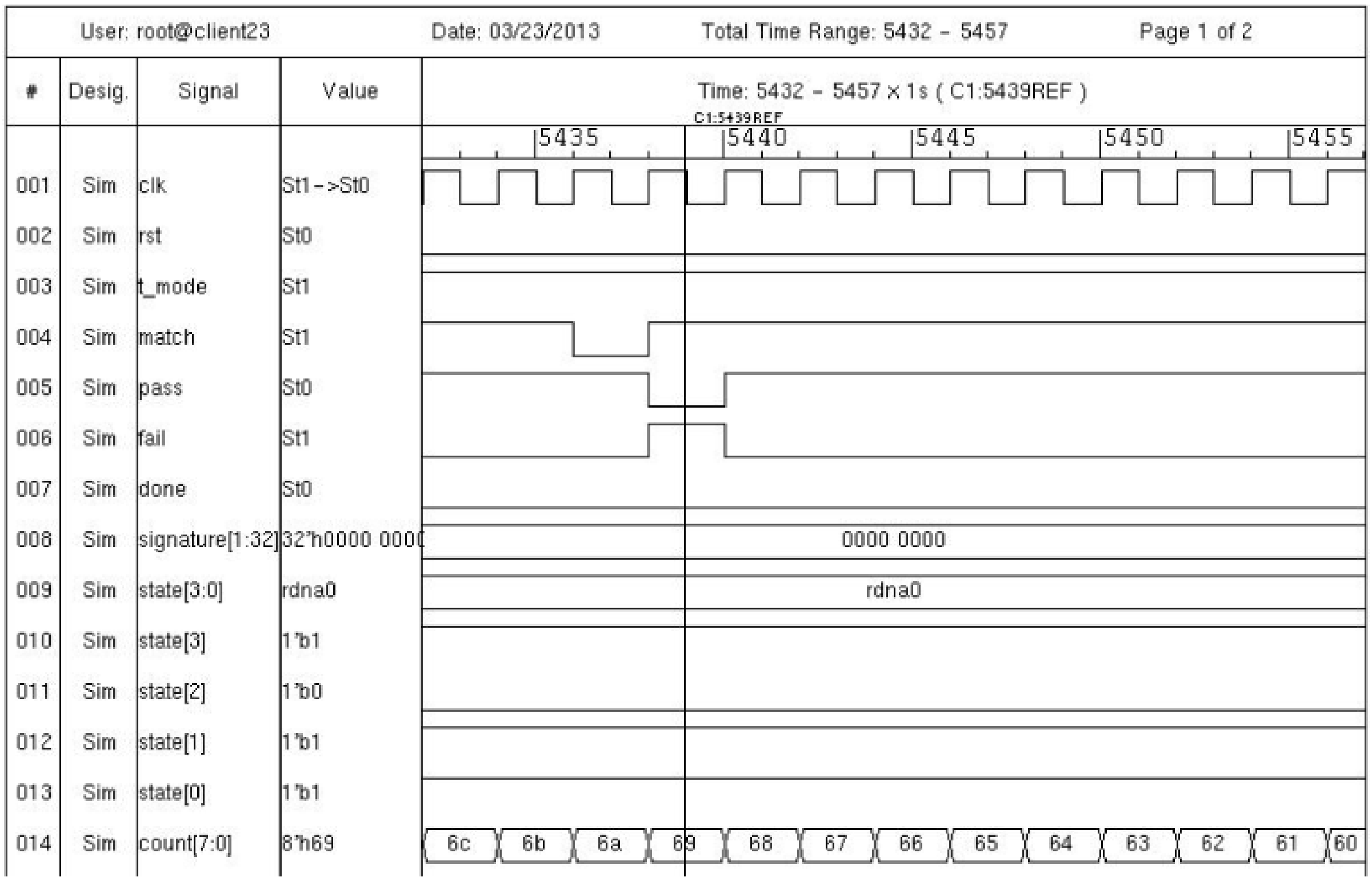

**Figure 5.5 Simulation output of MBIST controller.**

If match signal value is high then it shows that the output pattern matches with signature. It means there is no defect in memory. Now if match becomes low at any time it shows the detection of fault in memory. After this, with in one clock pulse, the fail signal will be asserted and simultaneously the pass signal will be deasserted.

# Chapter 6

# Functional Verification of MBIST controller

The functional verification of RTL model of MBIST controller is done using two different approaches one is the traditional way to verify the RTL model and another is assertion based dynamic verification.

The traditional functional verification used some test cases and based on those test cases the functionality of design is verified. But to know whether design has performed correct functionality or not, it has to look or watch signal to signal in the wave window generated by HDL simulator.

The simulation waveforms and coverage reports of traditional verification and assertion based dynamic verification is shown in chapter 8.

## 6.1 Assertion Based Dynamic Verification with Verification Plan

System Verilog Assertions (SVA), as a set of System Verilog language are embedded [15] into the design code of MBIST controller enabling the simulator to check the assertions during the dynamic verification of MBIST controller. A cross checks between intended and actual design of MBIST controller is provided by assertions. Assertions have been written from the specification/properties of the design, from design document and from the understanding of the FSM design of MBST controller.  Increased number of successful assertions increases the functional coverage of design which ensures that the design's functionality has been verified exhaustively not only at top level but also internally.

Now the verification plan for MBIST controller follows the following steps.

### 6.1.1 General Information about MBIST Controller as DUV

MBIST controller based on modified March C algorithm is implemented for design under verification. It controls the MBIST testing operation on SRAM. Controller is design by keeping in mind to test 32 bit single port SRAM under test. The controller design comprises the total 13 states where chosen March algorithm has 11 states and two extra states for controller idle and done state. There are some primary inputs and outputs of controller. The detail information of MBIST controller is presented in implementation of MBIST controller.

**6.1.2 MBIST Controller Interface's Verification Requirements at Block Level**

MBIST controller is considered as a block box [3] shown in Figure6.1. Here, only the interfacing signals between controller and other MBIST circuitry such as pattern generator, read/write generator, address generator and signature analyser are considered. Interface signals are verified for the timing requirement mainly by writing the assertions and by checking whether the test cases are invoking the interface signals or not.

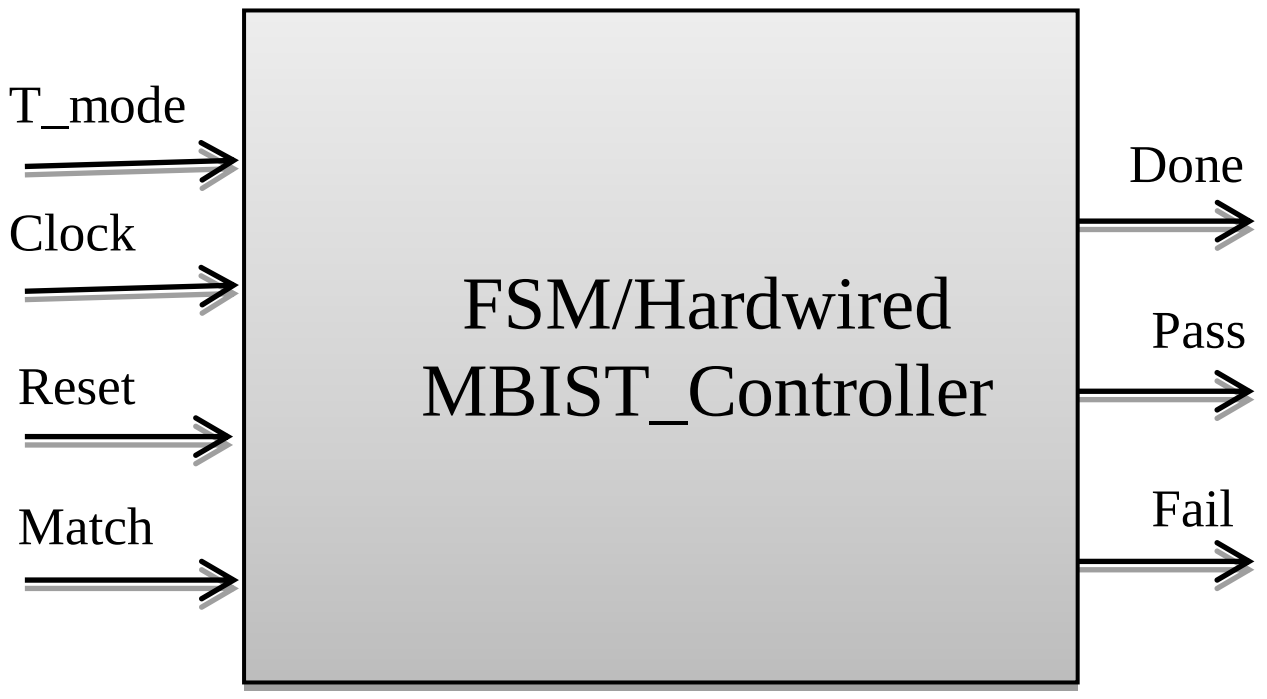

**Figure 6.1 MBIST controller as black box.**

**6.1.3 MBIST Controller Internal Operations and Functional Spots**

MBIST controller is the very crucial part of MBIST architecture. It requires great effort to verify its functionality because if something goes wrong with controller, it affects MBIST adversely. Here controller is implemented as a FSM, now checking whether the FSM works exactly the same way as intended in design specification is the way to verify the controller. The FSM's transitions and decisions are main internal functional spot to verify [12] in FSM of MBIST controller. Transitions and Timing requirements are verified using System Verilog assertions.

### 6.1.4 Assertions

Here, plain language requirements of MBIST controller are converted into formal properties using System Verilog Assertions. It would be mandatory to check the desirable requirements for the design by using assertions in real verification projects. A function with more than two logic events is verified using a SVA keyword ‘sequence’. A simple logic change is an event, can also be change in expression or Boolean sub expression. A SVA key word ‘property’ is used to verify the complex function of controller and it also includes the sequences for some complex functions [2]. These defined properties are verified during dynamic verification and checked by using a key word ‘assert’.

### 6.1.5 Code Coverage Goal

Before code coverage analysis and to know the completeness of verification, the code coverage goal has to be set. Code coverage involves different types of coverage metrics which their own coverage goals. Here, line coverage, statement coverage, block coverage, branch coverage, path coverage, conditional coverage, event coverage and FSM coverage are planned to analyze against their planned coverage goal. The coverage goals are set to be 100% as shown in Table 6.1. If any design portion or code is not covered then it is explained.

**Table 6.1 Code Coverage Goal**

| Code Coverage | Coverage Goal |
|---|---|
| Line Coverage | 100% explained |
| Statement Coverage | 100% explained |
| Block Coverage | 100% explained |
| Branch Coverage | 100% explained |
| Path Coverage | 100% explained |
| Conditional Coverage | 100% explained |
| Event Coverage | 100% explained |
| FSM Coverage | 100% explained |

### 6.1.6 Functional Coverage Goal

The functional coverage for all assertion, toggle and FSM transitions & sequencing are planned. The assertions are categorized based on two functional levels one at interface and other inside the MBIST controller. Firstly the assertions are written to check the all interface requirements. Others will be checking the internal functionality requirements of MBIST controller. The functional coverage is planned by using the cover property for all assertions. The functional coverage goal for all of them is planned 100 % as shown in Table 6.2.

**Table 6.2 Functional Coverage Goal**

| Functional Coverage | Coverage Goal |
|---|---|
| All Cover Properties | 100 % |
| Toggle | 100 % |
| FSM Transitions | 100 % |
| FSM Sequencing | 100 % |

## 6.2 Implementation of Verification Plan

Implementation of Verification Plan for MBIST controller's assertion based verification is presented in this section. It comprises two main subsections.

### 6.2.1 Four Steps towards Assertion Based Verification of MBIST Controller

Assertion Based Verification addresses all the challenges faced during only simulation based validation of design without the use of assertions. Here, assertions are defined for MBIST controller with complete assertion's ontology [11] which includes the information about assertion under construction such as class, type, name, expression, message, severity level, condition and snippet. The SVAs are written for MBIST controller to check the required specifications and functional properties according to verification plan. Based on the verification requirements, assertions for MBIST controller design are categorized only on the basis of class and other ontological information would be covered with this.

A four step approach is used for assertions based functional verification of MBIST controller.

**i) At Interface Boundary of MBIST Controller:** It is a block level based approach to write the assertions for MBIST controller. It is mentioned in section 2 that MBIST controller interacts with several blocks in MBIST architecture and other than MBIST too. SVAs are written here to check whether the controller interface/boundary signals are according to the required controller's specification or not. Here, assertions mainly check the timing requirements among the interface signals [12]. The timing requirements for interface signals to function correctly is shown in Table 6.3. And to check these interface requirements the SVAs examples are shown in Figure 5.4.

**Table 6.3 Verification requirements for the interface signals.**

| **Interface Signals** | **Interface Signals their features** | |
|---|---|---|
| | Signals | Expected Features |
| 1. | Reset, T_mode | T_mode=1 and reset=0 and its opposite must be synchronous |
| 2. | T_mode, en | en=1 with in 1clock cycle after t_mode is asserted |
| 3. | Match, Pass | Pass=1 with in 1clock cycle after match is asserted |
| 4. | Match, Fail | Fail=0 with in 1clock cycle after match=0 |

**ii) Internal Functional Spots of MBIST Controller:** MBIST controller is the control block of MBIST which is implemented as a FSM. The main verification requirements of FSM are defined as the transitions among the states, sequencing and timing requirements. For the every March operation shown in Figure 5.2, there is a different state in controller's FSM design. Some of the state transitions are shown in Table 6.4.

Now here, assertions are written according to verification plan and classified to verify the legal state transitions, correct sequencing, timing requirements and functionality in each state.

**Table 6.4State transitions those need to be verified.**

| State Transitions | Next States | | | | | | | | | | | | |
|---|---|---|---|---|---|---|---|---|---|---|---|---|---|
| | S_idle | Wdn0 | Rup0 | Wup1 | Rdn1 | Wdna0 | Pause | Rdn0 | Wdn1 | Rup1 | Wup0 | Rdna0 | S_done |
| 0 (S_idle) | | ✓ | | | | | | | | | | | |
| 1 (Wdn0) | ✓ | | ✓ | | | | | | | | | | |
| 2 (Rup0) | ✓ | | | ✓ | | | | | | | | | |
| 3 (Wup1) | ✓ | | | | ✓ | | | | | | | | |
| 4 (Rdn1) | ✓ | | | | | ✓ | | | | | | | |
| 5 (Wdna0) | ✓ | | | | | | ✓ | | | | | | |
| 6 (Pause) | ✓ | | | | | | | ✓ | | | | | |
| 7 (Rdn0) | ✓ | | | | | | | | ✓ | | | | |
| 8 (Wdn1) | ✓ | | | | | | | | | ✓ | | | |
| 9 (Rup1) | ✓ | | | | | | | | | | ✓ | | |
| 10 (Wup0) | ✓ | | | | | | | | | | | ✓ | |
| 11(Rdna0) | ✓ | | | | | | | | | | | | ✓ |
| 12(S_done) | ✓ | | | | | | | | | | | | |

Each check mark in Table 6.4 shows that the state transitions between those state and next state must be verified. R, W, up, dn stands for read operation on to MUT, write operation, upward marching means from min address to the max address of MUT and the down marching means max address to min address, respectively. The SVAs written for all required transitions to check whether the transitions are happening in correct manner or applied test cases are exercising all required transitions or not. Two SVAs examples are shown in Figure 6.4. All used assertions are tabled with their explanation in Appendix-A.

**Table 6.5Functional behavior of MBIST controller in different states need to be verified.**

| State's Transitions | States and their internal features | |
|---|---|---|
| | States | Expected features |
| 1. | S_idle | Control signal en=0 to other MBISTcircuitry, Done=0; |
| 2. | Wdno | Initially counter must consists count value (count=c_max), lastly count=c_min, The read/write control signal rw=0 and pattern generate control signal g_patt=0; en=1; Done=0; |
| 3. | Rup0 | Initially count value (Count =c_min), lastly count=c_max, rw=1, g_patt= must hold the previous state value, en=1; Done=0; |
| 4. | Wup1 | Initially count=c_min, lastly count=c_max, rw=0, g_patt=1; en=1; Done=0; |
| 5. | Rdn1 | Initially count=c_max, lastly count=c_min, rw=1, g_patt=previous state value, en=1; Done=0; |
| 6. | Wdn0a | Initially count=c_max, lastly count=c_min, rw=0, g_patt=0, en =1; Done=0; |
| 7. | Pause | Initially count=c_min, lastly count=c_max, all other control signal will same as in previous state. It holds the further operation for a particular interval of time, Done=0; |
| 8. | Rdn0 | Initially count=c_max, lastly count=c_min, rw=1, g_patt=previous state value, en=1; Done=0; |
| 9. | Wdn1 | Initially count=c_max, lastly count= c_min, rw=0, g_patt=1,en=1; Done=0; |
| 10. | Rup1 | Initially count=c_min, lastly count=c_max, rw=1, g_patt=previous state value, en=1; Done=0; |
| 11. | Wup0 | Initially count=c_min, lastly count=c_max, rw=0, g_patt=0,en=1; |
| 12. | Rdn0a | Initially count=c_max, lastly count=c_min, rw=1, g_patt=previous state value, en=1, Done=0; |
| 13. | S_done | count=c_min, Done=1; no further march test. |

Table 6.5 shows the functionality of MBIST controller in each state which needs to be verified. In every state, controller asserts or does not assert the some control signal which controls the other MBIST circuitry according to the testing algorithm shown in Figure 5.2. Figure 6.4 shows the SVAs examples which are embedded into MBIST controller's HDL code to verify the features defined in the Table 6.5. The Fig 6.5 represents their result of success.

**iii) Formalized the properties with SVA:** Here we converted plain language requirements into formal properties using System Verilog Assertions. A function with more than two logic events is verified using a SVA keyword 'sequence'. A simple logic change is an event, can also be change in expression or Boolean sub expression. A SVA key word 'property' is used to verify the complex function of controller and it also includes the sequences for some complex functions [2]. These defined properties are verified during dynamic verification and checked by using a key word 'assert'. Figure 6.3shows some complete system Verilog assertions and cover properties for functional requirement at interface boundary of MBIST controller.

```
1) property check_en;
@(posedge clk) disable iff(!t_mode) (t_mode)|->en; endproperty
Ap5:assert property(check_en)
$display("enable rise is ok");
else
$warning("enable gets delay");
Cp5:cover property (check_en);

2) property rose_t_mode;
@(posedge clk) disable iff(rst) (t_mode); endproperty
Ap8: assert property (rose_t_mode)
$display("t_mode and rst are synchronized");
else
$error("t_mode and rst are not synchronized");
Cp8:cover property (rose_t_mode);
```

**Figure 6.2 Assertions for interface features.**

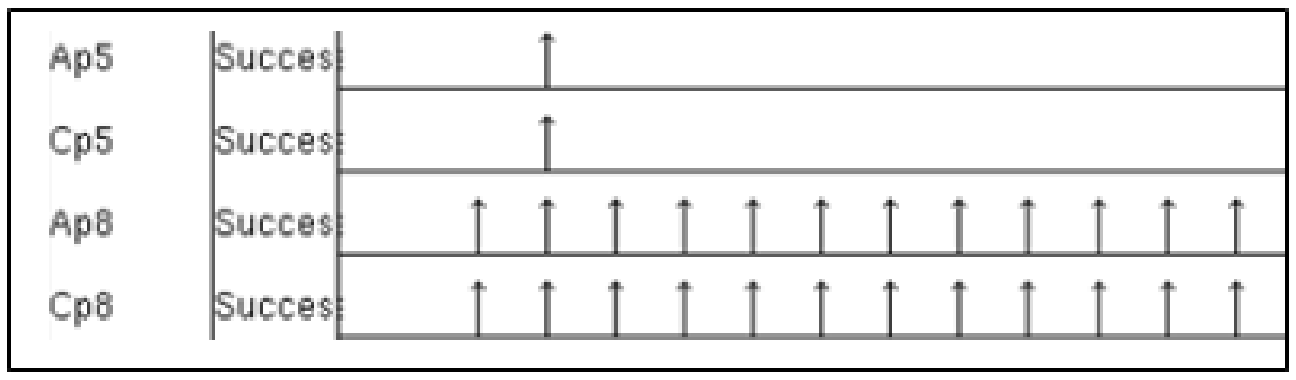

**Figure 6.3 Result of SVAs from Figure6.2 during simulation.**

The upside arrows for asserted properties Ap5 & Ap8 show the success of assertion and for cover properties Cp5 & Cp8 shows that the functionality defined in assertion is exercised by the test cases. And no need to add more test cases to check that particular functionality of design. The success of Ap5 shown in Figure 6.3 indicate that the control signal en to the address generator and analyzer gets high value within 1 clock cycle after getting high value at t_mode pin of MBIST controller.

```
1)sequence check_rup0;
(state==rup0);endsequence   //Define Sequence
Sequence check_idle;
((rst==1)&&(state==s_idle));endsequence
property rup0_idle;           //Define Property
@(posedge clk) check_rup0|=>check_idle; endproperty
Ap_rup0_idle:assert property(rup0_idle); //Define Check Point
Cp_rup0_idle:cover property(rup0_idle); //Define Cover Spot

2) sequence state_wdn0;
(state==wdn0) && (count==c_min); endsequence
property rup0_tran;
@(posedge clk) stat_wdn0|-> ##1 state==rup0; endproperty
Ap13:assert property(rup0_tran)
$display ("rup0 state transition is legal");
else
$warning ("rup0 state transition is illegal");
Cp13:cover property(rup0_tran);
```

**Figure 6.4 SVAs for state transitions verification.**

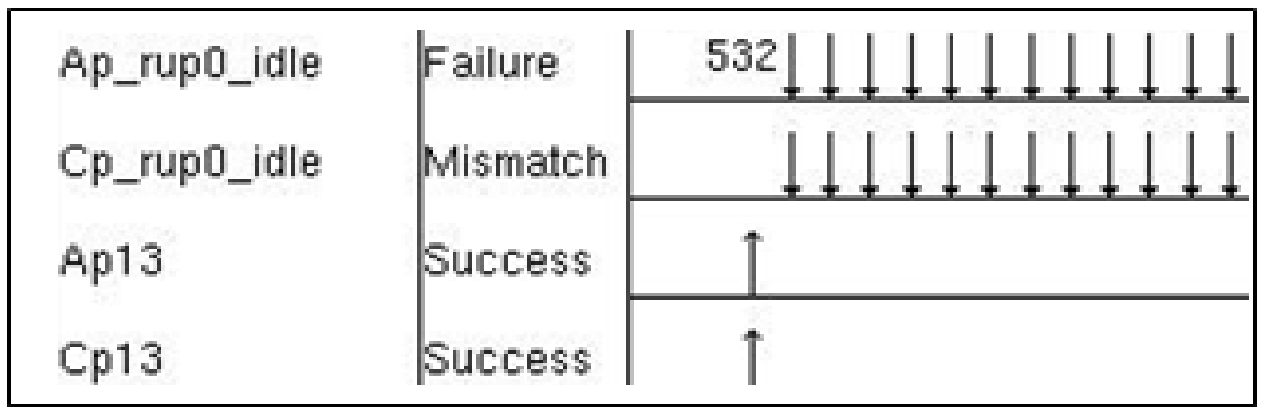

**Figure 6.5Result of SVAs from Figure 6.4 during simulation.**

The results of SVAs checkers written in Figure 6.4 are shown in Figure 6.5. These SVAs are used to verify the transition from state wdn0 to state rup0 succeeded as Ap13 & Cp13 and to verify the transition between state rup0 and state s_idle got fail and cover property gave the mismatch results. The failures of SVA checker rup0_idle indicate that there is still untested area in design or there is a loop hole between the intended and designed FSM of MBIST controller. But at the same time the property coverage is also not true. At first, the test cases are improved to exercise that particular transition.

```
1)  sequence s1;
state ==wdn0; endsequence
sequence s2;
! (rw||g_patt); endsequence
//Define sequence
property f_wdn0;
@ (posedge clk) s1 |->s2; endproperty
//Define property
Ap34: assert property (f_wdn0);
//Define check point
Cp34: cover property (f_wdn0);
//Define coverage spot
2)  property rdna0_done;
@ (posedge clk) ((state==rdna0)|=> ((state==done)
&& (count==c_max))); endproperty
//Define a property
Ap31: assert property (rdna0_done);
Cp31: cover property (rdna0_done);
//Define coverage spot
```

**Figure 6.6 Examples of system verilog assertions for MBIST controller.**

In Figure 6.6, there are two MBIST controller properties are defined as System Verilog Assertions because it includes the both assertions and functional coverage. First property f_wdn0 consists two sequences s1 and s2. This property checks the signal g_patt and rw signal values during the state wdn0. Signal g_patt is the input to pattern generator and signal rw is the input to the read/write generator. In Figure 6.6, the property rdna0_done is written in the form of SVA to verify the legal transition from state rdna0 to state done.

**iv)Functional coverage:** Here we defined functional coverage for all the formalized properties of MBIST controller in above step. Mainly assertions have two functions one as checker and another is as functional coverage [15]. It is illustrated in Figure 6.6, that functional spot is also defined as coverage spot. Figure 6.7 shows run time log file obtain from verification tool VCS which contains detailed assertion message. Now the errors in the design are found and debugged easily. Level of severity has been defined based on the effect of property failure in MBIST controller design. By default severity level of an assertion is ‘0’ and severity of assertion is defines as ‘error’ at the failure of assertion in Synopsys VCS verification tool. System Verilog has four severity levels $error, $warning, $info and $fatal [6]. Severity level of $fatal is so much that it terminates the simulation immediately at failure of assertion.

```
"nc_mbistcontroller.sv", 157: tb_bist_cont.uut.Ap6: started at 6442s failed at 6442s
        Offending '$stable(state)'
Warning: "nc_mbistcontroller.sv", 157: tb_bist_cont.uut.Ap6: at time 6442
state transition is not stable with rst
property is successed
```

**Figure 6.7 Portions of run time log file for assertion Ap6.**

By examine the run time log file shown in Figure 6.7, it can be easily found out which assertion failed and at what location.

A run time log file consists the complete information about every assertion. Here in Figure 6.7, it is very clearly the location (at line 157) of assertion Ap6 and name of module (nc_mbistcontroller.sv) in which Ap6 had been embedded. Assertion Ap6 failed at time 6442 and with property which could not hold at this time. The severity attached with this assertion is ‘warning’ with a message which is optional to include with the failure of assertion. After this time the property succeeded.

# Chapter 7

# Simulation Setup

This chapter describes the simulation setup [30] of EDA tool Synopsys-VCS® during MBIST controller RTL implementation and its assertion based functional verification.

## 7.1 Synopsys-VCS® Setup for Complete Simulation

VCS software's setup is done for the compilation, simulation and debugging of the test-bench. The software is only capable to run only on the Red Hat Linux running system. The software is installed at Linux running server. The communication with the server to access the license file is done through a SSH client.

### 7.1.1 Test-bench Compilation

For the test-bench compilation, the vcs command is used and suitable switches are added for required operations with this. The following command is used to compile the test-bench**.**

```
Vcs –Mupdate –lca –sverilog –cm line+cond+path+branch+tgl+fsm+assert –cm_noconst -cm_linecontassign –cm_condallops+anywidth+event –cm_fsmopt sequence –PP –debug_all *.sv –l compile.log
```

The compile time switches and their description is listed in Table 7.1.

**Table 7.1 Switches in compilation command and their description.**

| Switch Name | Switch Description |
|---|---|
| -Mupdate | To compile only modified module from last compilation |
| -lca | Limited Customer Availability to use extra added features |
| -sverilog | To compile system Verilog source files |
| -cm | To specify type of coverage |
| -cm assert | Compile for System Verilog assertions |
| -cm_noconst | Ignore constant expression and unreachable statements |
| -cm_linecontassign | To monitor continuous assignment |
| -cm_condallops+anywidth+event | Monitor non-logical operators of any width and always block sensitivity expressions for conditional coverage |
| -cm_fsmopt sequence | To cover the sequences in FSMs |
| -PP | Enables the debuging |
| -debug | Compile for debug in DVE |
| -debug_all | Compile for debug in DVE with linestepping features to debug the design |
| -l compile.log | To save compiler message in log file during compilation |

The + sign in cm switch enables the all listed coverage simultaneously. During compilation some object files are generated which are linked to a executable file simv (by default)in working directory. This file name can be changed by using a switch. This executable file simv is used by tool for simulation.

### 7.1.2 Simulation

The executable files generated during compilation are used to start simulation. During simulation, switches are also used to do required operations. The following simv command is used to simulation.

```
simv -cm line+cond+path+branch+tgl+fsm+assert -l run.log-assert -gui
```

The Table 7.2 contains the switches used in simulation command with their descriptions.

**Table 7.2 Simulation time switches and their description**

| Switch Name | Switch Description |
|---|---|
| -cm line+…+assert | For Code coverage collection |
| -assert | To enable assertion setting |
| -l run.log | To Save simulation time log file |
| -gui | To enable the visual debugging interface- DVE |

### 7.1.3 Debugging

To ensure the correct functionality of MBIST controller design, the complete code was tested and debugged. For simulation and debugging purpose, Synopsys's VCS-DVE is used. DVE provide the graphical user interface by showing signal waveforms for debugging. Where, DVE stands for Discovery Visualization Environment. The detail description of Synopsys's VCS is described in section 4.2.

### 7.1.4 Code, Functional, Assertion and Property Coverage Reporting

The Universal Report Generator (URG) tool is used to generate the functional and assertion coverage reports. The cmView tool is used to generate the code coverage reports.

The following command is used for code coverage report generation.

```
vcs –cm_pp –cm_report summary
```

After that the code coverage report is saved in simv.cm directory in working directory. This generates the combine coverage report for all code coverage metrics as well as separate coverage metrics. During code coverage generation the assertion coverage report is generated separately in the same simv.cm directory.

The following command is used for the functional coverage report generation.

```
urg –dirsimv.vdb –format text
```

The generated report can be found in simv.vdb directory [30]. By default the generated report are saved in HTML format but the switch `–format text` is used to save the reports in text format in the same simv.vdb directory. The assertion property coverage report is saved during functional coverage generation while assertion coverage report is generated and saved during code coverage. The VCS tool generates the assertion report containing the number of attempts, success, failures and incompleteness during the simulation. While to monitor the sequences and other behavioral aspects of design, property coverage is generated by the tool. For this a cover property statement is used. In assertion property coverage reports, simulator saved the number of time a property holds or fails.

The complete simulation and coverage reports are shown in chapter 8.cmview is used to get the code coverage in both batch and gui mode. In batch mode, cmview saves the report in text format whereas in gui it shows graphically or pictorially. The different results and reports are analyzed in chapter 8.

# Chapter 8

# Results and Analysis of Coverage Results

This chapter will discuss the MBIST controller's design implementation and coverage results achieved during the simulations. Coverage results will comprises the code coverage, functional coverage and their various types of coverage metrics. Here, the analysis of code coverage metrics and functional coverage metrics are discussed in details for MBIST controller design.

### 8.1 MBIST Controller Simulation Waveform Description

The simulation output in Figure 8.1 shows input and output waveforms of MBIST controller. The '1' value of pass or fail signal indicates the success and failure of corresponding applied data pattern on the memory. When the match signal gets '0' value in the next clock cycle, MBIST controller assert the fail high and low to the pass signal. Again when match goes high the controller assert pass signal high and low to fail. Whenever rst (reset) signal goes high, the state immediately switches to the idle state. After attaining idle state, there is no transition in the state of controller and state gets stable until rst gets low value again. Controller starts transitions state to next state only when rst is low and t_mode signal gets high value.

Simulation output shown in Figure 8.1 is only based on some test cases and provides the functionality check but not completely. With such functional checking, there is a possibility to leave some corner cases untested. And there is possibility of bugs in such untested corner cases. These corner cases are tested by using system Verilog assertion for design's functional properties and design specification as shown in Figure 8.2. The functional coverage score without assertion and using all possible test cases is shown in Figure 8.3.After that only those test cases are used which have the maximum possibility to provide 100 % assertion coverage because the now functional verification of MBIST controller is depend on assertion checking.

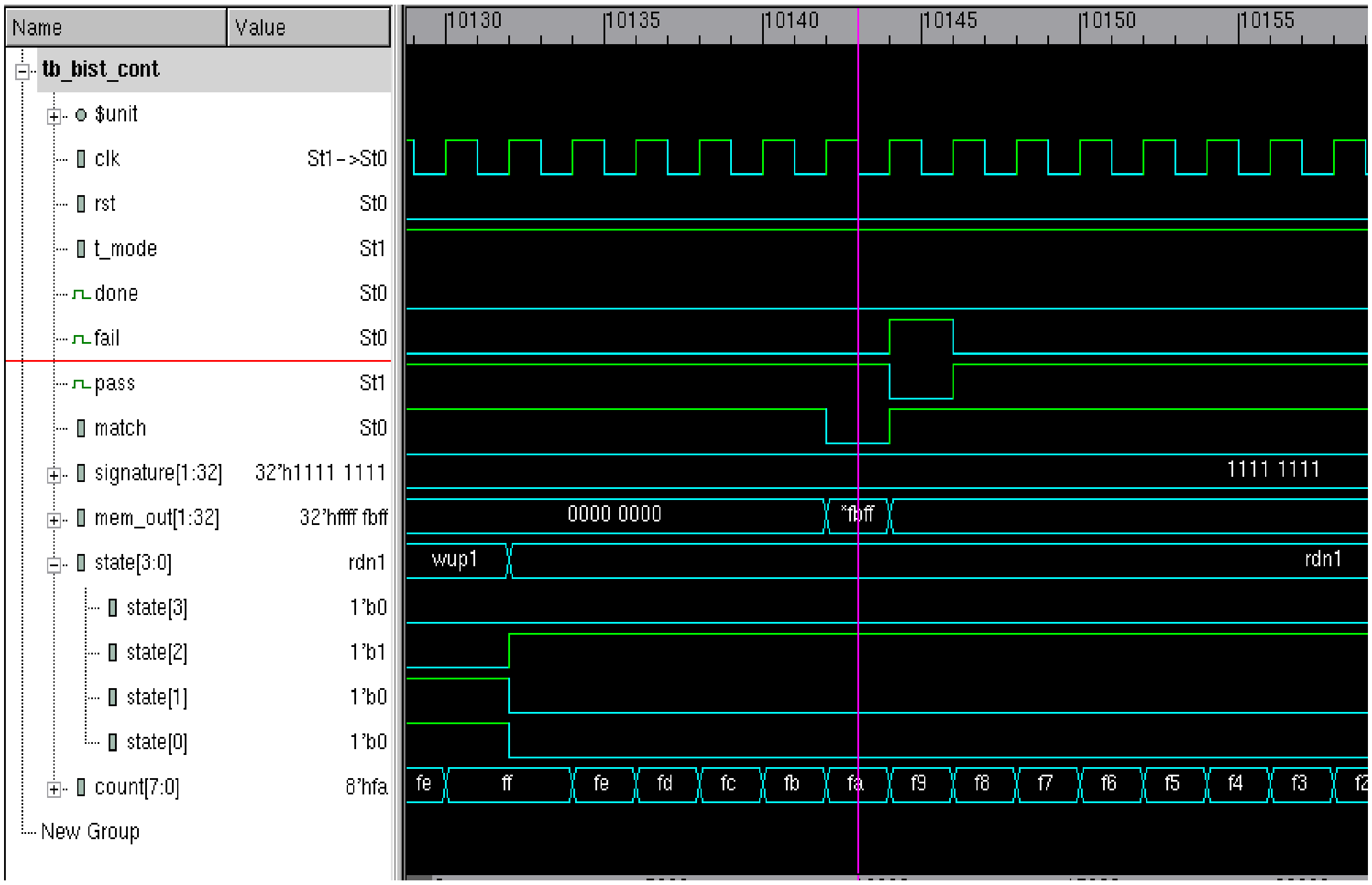

**Figure 8.1Simulation waveform of MBIST controller design without assertions.**

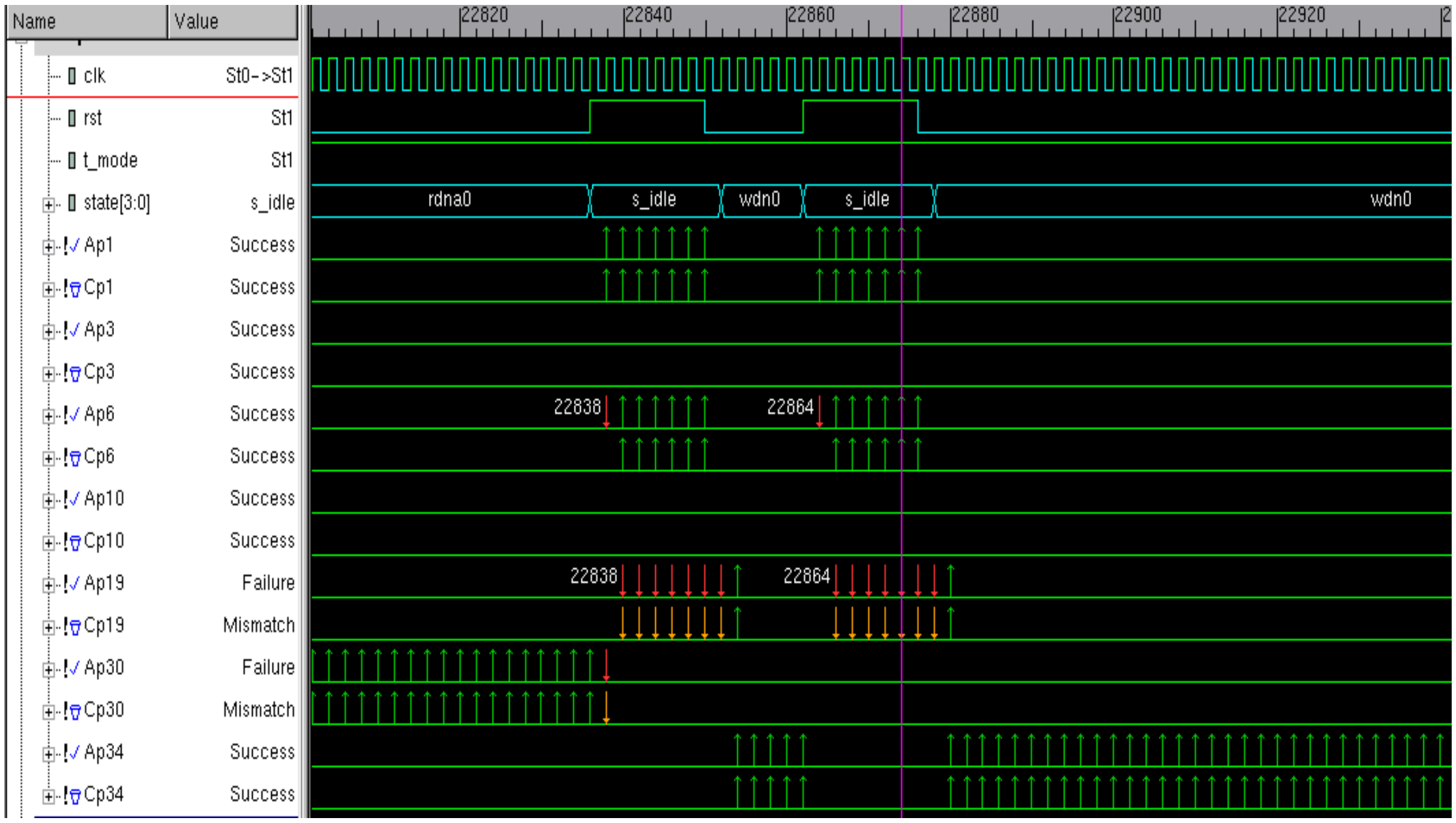

**Figure 8.2 Simulation waveform of MBIST controller with assertions.**

The assertion Ap6 is defined to check whether the state is stabilized in next clock cycle or not when rst signal gets a value ‘1. In Figure 8.2, the assertion Ap6 gets failed at time 22838 and 22864. These are times of positive edges of clock cycles just after clocks at which state transitions took place to the idle state. In the first clock cycle, state gets change to the idle state and waits there for the next clock to check whether the state gets stabilized or not. That’s why at time instances 22838 and 22864 assertion gets failed after that it succeeded. The row of Cp6 in Figure 6.7 has all up arrows means during the complete verifying time of asserted property, property gets covered.

The property Cp30 is that state will be remains in the state rdna0 until the address counter value is not equal to minimum value but in the mid of that rst gets a toggle from low to high. Due to that state change to state idle and the property written for state rdna0 gets mismatch at next clock because at this time no rdna0 state exits.

## 8.2 Verification Results

Verification results for the MBIST controller include the all code coverage metrics scores and functional coverage score.

### 8.2.1 Functional Coverage Results

Functional coverage results for MBIST controller verification include the total achieved functional coverage score, assertion coverage score, detailed assertions report and detailed cover property report.

The total functional coverage score without assertions is shown in Figure 8.3 and with assertions coverage score is shown in Figure 8.4. Detailed assertion and cover property reports are listed in the appendix-B.

```
-----------------------------------------------------------------------------
Total Coverage Summary
SCORE   LINE   COND   TOGGLE   FSM    BRANCH
 97.06   99.80  100.00  90.75    100.00    94.74
-----------------------------------------------------------------------------
Hierarchical coverage data for top-level instances
SCORE  LINE   COND   TOGGLE   FSM    BRANCH NAME
97.06   99.80  100.00   90.75    100.00   94.74    tb_bist_cont
```

**Figure8.3 Total functional coverage score without assertions.**

```
Version: D-2009.12
Command line: urg -dir simv.vdb simv.cm -format text
--------------------------------------------------------------------------
Total Coverage Summary
SCORE  LINE   COND   TOGGLE   FSM   BRANCH ASSERT
  96.50  99.35   100.00    90.17    100.00   89.47    100.00
--------------------------------------------------------------------------
Hierarchical coverage data for top-level instances
SCORE   LINE   COND   TOGGLE  FSM   BRANCH  ASSERT   NAME
 96.50    99.35    100.00    90.17    100.00    89.47    100.00   tb_bist_cont
```

**Figure 8.4 Total function coverage score with assertions.**

According to verification plan for MBIST controller design all required internal andinterfacial functional features are exercised using SVAs which helped to apply the efficient or directed test cases to the controller. Table 10 shows the directed test cases towards the desired functionality check with total functional coverage of MBIST controller. 37 out of 53 assertion got success with only 6 directed test cases. Based on the failed assertions, the design bugs are removed and the test cases are improved to get the 100 percent success of the assertions. After applying 25 directed test cases 100 percent assertion coverage is obtained, as shown in Figure 8.4 and 96.50 percent functional coverage as shown in Table 8.1. The main advantage using directed cased in ABV of MBIST controller is that it reduces the number of test cases from 88 to 25 with only .46 % decrement in total functional coverage score but with 100 % assertion coverage.

**Table 8.1Functional coverage score with different number of test cases and different number of assertions.**

| **No. of Directed Testcases** | **No. of SVAs in Verification Plan** | **No. of successful assertions** | **Total Functional Coverage Score** |
|---|---|---|---|
| 6 | 53 | 37 | 77.86 % |
| 25 | 53 | 53 | 96.50 % |

8.2.2 Code Coverage Results

The Figure 8.5 shows the combined report for the various code coverage metrics discussed in the section 3.4.1. In Table 8.2, the targeted and achieved coverage scores are presented for various coverage metrics.

```
+------------------------------+--------------------+--------------------+--------------------+
|module/instance               |        line        |     conditional    |       toggle       |
|name                          | !cov  |  tot  |  % | !cov |  tot  |  %  | !cov |  tot  |  %  |
+------------------------------+--------------------+--------------------+--------------------+
|* TOTAL_MODULEDEF             |      2|   1020| 99.80|     0|     24|100.00|    16|    173| 90.75|
|* mbist_cont                  |      2|    114| 98.25|     0|     24|100.00|    16|    102| 84.31|
|* tb_bist_cont                |      0|    906|100.00|    --|     --|    --|     0|     71|100.00||
|* \$unit::nc_mbistcontroller.sv::tb3_mbist_controller.sv   |     --|     --|    --|     --|     --|
+------------------------------+--------------------+--------------------+--------------------+
|module/instance               |        state       |     transition     |      sequences     |
|name                          | !cov  |  tot  |  % | !cov |  tot  |  %  | !cov |  tot  |  %  |
+------------------------------+--------------------+--------------------+--------------------+
|* TOTAL_MODULEDEF             |      0|     13|100.00|     0|     24|100.00|     1|    841| 99.88|
|* mbist_cont                  |      0|     13|100.00|     0|     24|100.00|     1|    841| 99.88|
|* tb_bist_cont                |     --|     --|    --|    --|     --|    --|    --|     --|    --|
|* \$unit::nc_mbistcontroller.sv::tb3_mbist_controller.sv   |     --|     --|    --|     --|     --|
+------------------------------+--------------------+--------------------+--------------------+
|module/instance               |        path        |       branch       |      assigntgl     |
|name                          | !cov  |  tot  |  % | !cov |  tot  |  %  | !cov |  tot  |  %  |
+------------------------------+--------------------+--------------------+--------------------+
|* TOTAL_MODULEDEF             |      7|     38| 81.58|     2|     38| 94.74|    --|     --|    --|
|* mbist_cont                  |      7|     38| 81.58|     2|     38| 94.74|    --|     --|    --|
|* tb_bist_cont                |     --|     --|    --|    --|     --|    --|    --|     --|    --|
|* \$unit::nc_mbistcontroller.sv::tb3_mbist_controller.sv   |     --|     --|    --|     --|     --|
```

**Figure8.5Coverage score for various code coverage metrics.**

**Table 8.2 Code coverage metric's and functional coverage metric's achieved score against targeted scores.**

| **Coverage Metrics** | **Targeted Goals (%)** | **Achieved Scores (%)** |
|---|---|---|
| Line | 100 | 99.80 |
| Conditional | 100 | 100 |
| Toggle | 100 | 90.75 |
| Path | 100 | 81.25 |
| Branch | 100 | 94.74 |
| State | 100 | 100 |
| State Transition | 100 | 100 |
| Sequencing | 100 | 99.88 |
| Assert | 100 | 100 |

The all achieved scores are analyzed and explain in the next section 8.3

All the results design and verification of MBIST controller are generated using Synopsys-VCS®.

## 8.3 Coverage Analysis of MBIST Controller Model

Coverage analysis makes functional verification easy and less time consuming and it also helps to know whether functional verification is done enough or not. Thats why coverage analysis plays important role to check the verification completeness of MBIST controller.

The controller's coverage analysis depends on two main coverages, one is code coverage and another is functional coverage. Code coverage is independent of functional coverage but it helps to obtain verification completeness of MBIST controller.

### 8.3.1 Code Coverage Analysis of MBIST Controller

Here eight different very useful coverage metrics are introduced that can be classified as code coverage metrics for MBIST controller. Code coverage metrics helped to identify which structure in HDL code of controller is exercised during simulation.

**i) Line Coverage:** Line coverage metric as Figure8.6 shows the result of exercised lines with respect to the total number of lines present in HDL code of MBIST controller**.**

| Module Name | Blks (%) | Blks | Stmnts (%) | Stmnts | Lines (%) | Lines |
|---|---|---|---|---|---|---|
| \$unit::nc_mbistcontroller.sv::tb3_mbist_controller.sv | | | | | | |
| | -- | 0/0 | -- | 0/0 | -- | 0/0 |
| tb_bist_cont | 99.80 | 508/509 | 99.80 | 1018/1020 | 99.75 | 400/401 |
| tb_bist_cont.uut | 98.08 | 51/52 | 98.25 | 112/114 | 97.06 | 33/34 |

**Figure 8.6 Line coverage score for MBIST controller.**

In Figure 8.6, the total line coverage score is around 98.5% for system verilog HDL code of MBIST controller . At the same time the Figure 8.7 shows the covered lines and uncovered line in HDL of controller. It helped to identify the cause of not covering the particular lines and shows the effectiveness of test suits.

**ii) Statement Coverage:**In this coverage analysis of MBIST controller, it only counts the executable statements. A line may contains more than one statements .For a conditional statement at line 36 in Figure8.7, there one countable statement state<= s_idle considered for *if* statement and multiple countable statements are considered for *else* statement [2]. Figure8.6 shows the total statement coverage score is around 99%.  In Figure8.8 at line 68, the red shaded portion shows the uncovered countable statements for else statement.

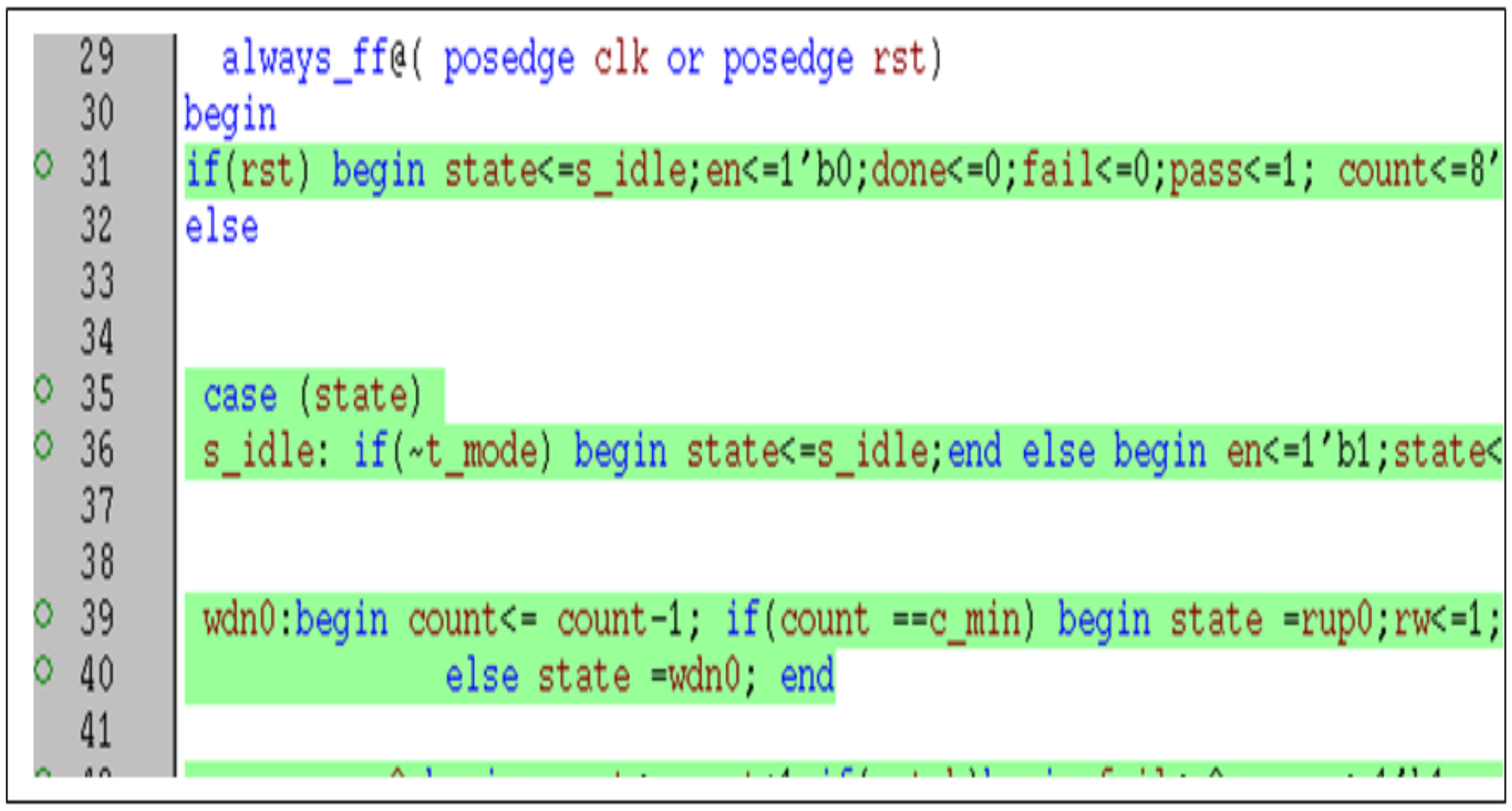

**Figure 8.7  Covered lines in the HDL code of MBIST controller.**

**iii) Block Coverage:** In Figure8.6, the block coverage score is shown as around 99% with total number of blocks extracted from the MBIST controller HDL code by Synopsys-VCS®.

**iv) Branch Coverage:**The total branch coverage score for designed controller is 94.74 shown in Figure8.8. It is very much possible that at initial attempts line coverage is around 100% and branch coverage is much less than of it. It shows that there are still untested cases.

| Module Name | Branches (%) | Branches |
|---|---|---|
| \$unit::nc_mbistcontroller.sv::tb3_mbist_controller.sv | -- | 0/0 |
| tb_bist_cont | 94.74 | 36/38 |
| tb_bist_cont.uut | 94.74 | 36/38 |

**Figure 8.8 Branch coverage score for MBIST controller**

As it measures the coverage of each branch present in *if* and *case* statements. It focused on the decision points or control statements that affect the control flow of the controller's HDL code execution. From a decision point, different branches originates in HDL such as form an *if*-statement, there are two branches one is for true and another is for false case. Decision coverage will report the evaluation in both true and false cases during simulation. For the *case* statement present in FSM HDL code of MBIST controller, decision coverage verified that each branch of the *case* statement, including the *default*. But the *default* state in the designed MBIST controller is idle state which is controlled by reset signal which is at high value by default. It showed uncovered default branch and one other at line 68 shown in Figure8.9.

```
64        wdn1:begin  count<=count-1; if(count==c_min)begin count<=c_min;rw<=1;
65             state=rup1;end else state= wdn1;end
66
67        rup1:begin  count<=count+1; if(match)begin pass<=1'b1;fail<=1'b0; end
68                             else begin fail<=1'b1; pass<=1'b0; end
69             if(count==c_max)begin state=wup0;g_patt<=0; rw<=0;count<=c_min;end
70             else state=rup1;end
71
72        wup0:begin count<=count+1; if(count==c_max)begin count<=c_max;rw<=1;
```

**Figure 8.9  Covered branches in HDL code of MBIST controller.**

The difference between line coverage and branch coverage of MBIST controller is due to untested default branch which is due to implied design of case statement in controller's HDL

Therefore, Branch coverage Metric is considered to be more complete than line coverage matric.

**v) Path Coverage:** The total path coverage score of MBIST controller is 81.58% as shown in Figure 8.10. If a branch is not covered in HDL code then it may stop the coverage of several paths through this branch and generates the sequencing error in the design that's why path coverage metric is more complete than the branch or decision coverage metrics. It is a very cumbersome because of very large number of paths in a design which makes 100 % score impractical for path coverage.

```
Module Name                           Path
                                      (%)
\$unit::nc_mbistcontroller.sv::tb3_mbist_controller.sv
                                      --
tb_bist_cont                          81.58
tb_bist_cont.uut                      81.58

//*******************************************************

//          Total Module Instance Coverage Summary

//                        TOTAL    COVERED    PERCENT
//    paths               38       31         81.58
```

**Figure 8.10 Path coverage score for MBIST controller**

In HDL code of controller, there are many if statements which generates the different branches and sequences. In other words, a control statement generates a different path in the HDL code. Path coverage is similar to decision coverage but it covers multiple sequential decisions. The branch of *case* statement (In the HDL code of MBIST controller) on line 67 followed by the *else* branch of *if* in Figure8.9 defines one path.

**vi) Conditional coverage:** The total conditional coverage score is 100% shown in Figure8.11.It means every possible case of conditions has been evaluated during simulation of HDL code of MBIST controller.

| Module Name | Logical (%) | Logical | Non-logical (%) | Non-logical | Events (%) | Events |
|---|---|---|---|---|---|---|
| \$unit::nc_mbistcontroller.sv::tb3_mbist_controller.sv | -- | 0/0 | -- | 0/0 | -- | 0/0 |
| tb_bist_cont | 100.00 | 22/22 | -- | 0/0 | 100.00 | 2/2 |
| tb_bist_cont.uut | 100.00 | 22/22 | -- | 0/0 | 100.00 | 2/2 |

**Figure 8.11 Conditional coverage scores for MBIST controller.**

It provides the coverage statistic for variables and expressions. It is shown in Figure8.12 as shaded portions (as evaluated expressions) how a conditional coverage metrics records the coverage by evaluating the variables or expressions. It is a very important and critical coverage metrics because it can find the errors in the conditional statements that cannot be easily found by any other coverage analysis.

```
wdn0:begin count<= count-1; if(count ==c_min) begin state =rup0;rw<
           else state =wdn0; end

      rup0:begin count<=count+1;if(match)begin fail<=0; pass<=1'b1
                                    else begin pass<=0;fail<=1'b1;
          if(count == c_max) begin state= wup1;g_patt<=1;rw<=0; c
             else state =rup0;end

      wup1:begin count<=count+1; if(count==c_max)begin state=rdn1;
          else  state= wup1; end
```

**Figure 8.12 Covered conditions in the HDL code of MBIST controller.**

**vii) Event Coverage:** Events are associated with the change of a signal. For example, as shown in the line 29 of Figure8.7, there are two event *always @ (posedgeclk or posedgerst)* which wait for the *clk* or *rst* signal changing from low to high. These two are the control events of complete FSM design of MBIST controller. It is shown in Figure8.11 that both events have been covered completely and event coverage score for the controller's HDL code is 100%. This coverage metrics is very useful when there are too much control events in the design.

**viii) FSM Coverage:** The FSM state traversed coverage score is 100% as shown in Figure8.13. The all 13 states have been traversed in HDL of MBIST controller written in system Verilog during simulation shown as shaded portion in Figure8.14.

Test Coverage Result: Total Coverage

| Module Name | States (%) | States | Trans. (%) | Trans. | Seq. (%) | Seq. |
|---|---|---|---|---|---|---|
| \$unit::nc_mbistcontroller.sv::tb3_mbist_controller.sv | | | | | | |
| | -- | 0/0 | -- | 0/0 | -- | 0/0 |
| tb_bist_cont | 100.00 | 13/13 | 100.00 | 24/24 | 99.88 | 840/841 |
| tb bist cont.uut | 100.00 | 13/13 | 100.00 | 24/24 | 99.88 | 840/841 |

**Figure 8.13 FSM coverage score for MBIST controller**.

In code coverage point of view, the FSM coverage metric cover number of traversed states in FSM design during the simulation. Here FSM coverage is defined as language-based code coverage for the MBIST controller HDL code just to show whether the all design states are traversed or not. It is most important coverage metric for presented MBIST controller design because it found out most of the design bug due to its closeness to the behavior of design space. The transitions and sequencing coverage of MBSIST controller's FSM is defined in the functional coverage.

The defining of FSM coverage metric was relatively more beneficial than other metric because it is MBIST controller design dependent. The main problem was to write the coverage directed tests so that all states, transition and sequences gets covered.

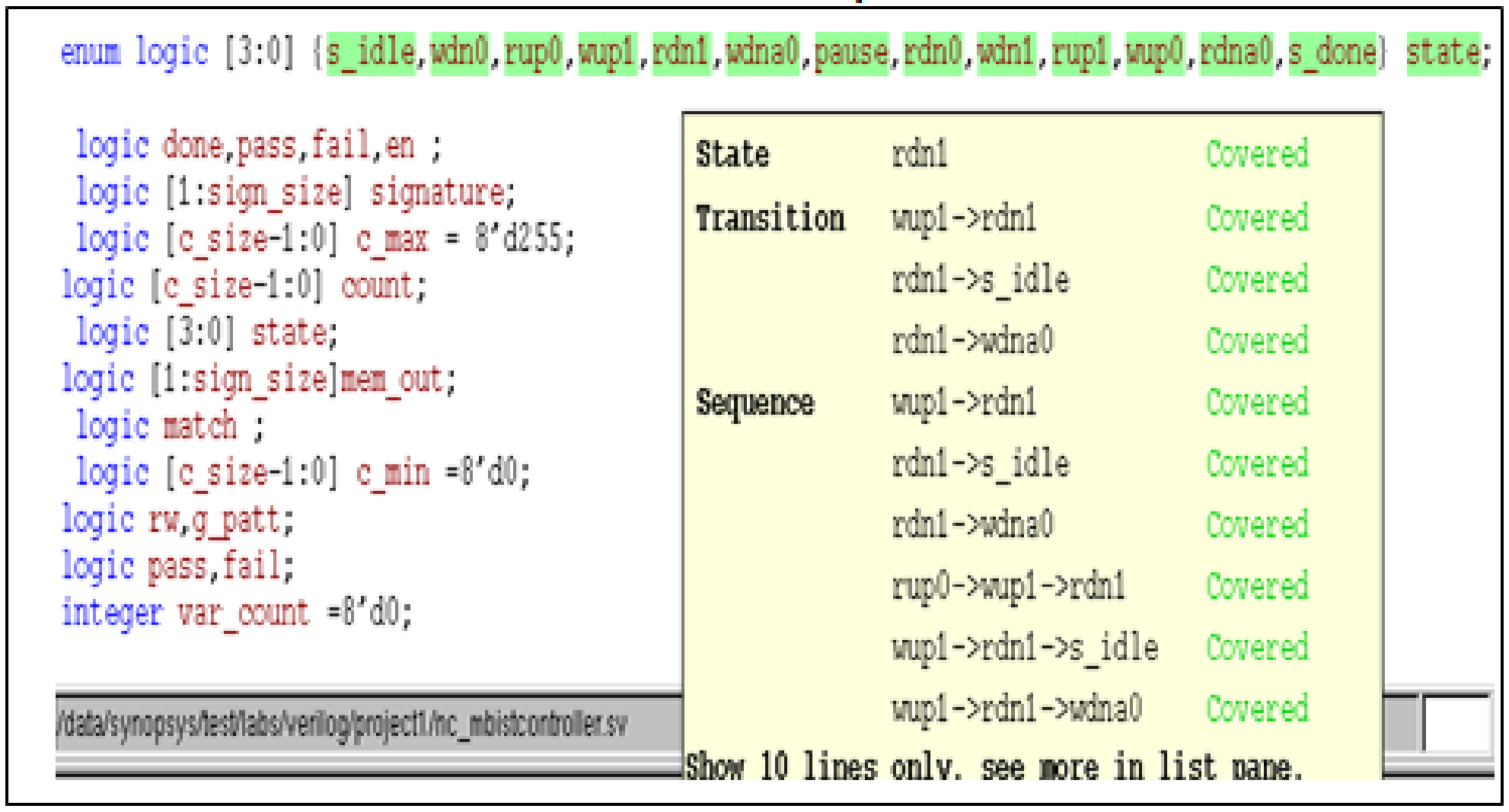

**Figure 8.14Covered states and transitions of MBIST controller FSM.**

### 8.3.2 Functional coverage Analysis of MBIST Controller

Here defined functional coverage metrics are corresponding to the designed FSM controller. Metrics defined in this category such as toggle coverage, Sequence and transition coverage of FSM and assertion coverage are related to computation performance of HDL code rather than its structure [24]. The main motive to define such metrics was to exercise each functional scenario of MBIST controller. During functional coverage, the coverage monitor looks for error in state transitions and event sequencing mainly in FSM designs. Assertions which interpret the MBIST controller hardware functionality in system Verilog language are considered the special case to monitor.

**i) Toggle Coverage:** The functional behavior of MBIST controlled is different for the toggling from 0 to 1 and from 1 to 0 of some logic bits. In Figure 8.15, detail toggle coverage report is shown with toggle score of net, register and logic bits. At same time the Figure8.16 shows that which logic signal bits did not toggled completely. In Figure 8.16 the red shaded signals shows that their bits are not changing polarity that's why not covered by coverage metric during simulation. But it is understood here that signal c_max and c_min in MBIST controller design are the minimum and maximum value of address counter which are constants values.

| // | TOTAL | COVERED | PERCENT |
|---|---|---|---|
| regs | 6 | 6 | 100.00 |
| reg bits | 68 | 68 | 100.00 |
| reg bits(0->1) | 68 | 68 | 100.00 |
| reg bits(1->0) | 68 | 68 | 100.00 |
| nets | 6 | 6 | 100.00 |
| net bits | 6 | 6 | 100.00 |
| net bits(0->1) | 6 | 6 | 100.00 |
| net bits(1->0) | 6 | 6 | 100.00 |
| logics | 13 | 11 | 84.62 |
| logic bits | 99 | 83 | 83.84 |
| logic bits(0->1) | 99 | 83 | 83.84 |
| logic bits(1->0) | 99 | 83 | 83.84 |

**Figure 8.15 Toggle coverage score for MBIST controller in detail.**

Toggle coverage analysis is proved very useful for MBIST controller functional verification because involvement of 32 bit read and write for the 32 bit wide SRAM. It is applied for structural and white box verification of controller to show that the control logic is working as intended.

```
 output done,pass,fail);

enum logic [3:0] {s_idle,wdn0,rup0,

 logic done,pass,fail,en ;
 logic [1:sign_size] signature;
 logic [c_size-1:0] c_max = 8'd255;
logic [c_size-1:0] count;
 logic [3:0] state;
logic [1:sign_size]mem_out;
 logic match ;
 logic [c_size-1:0] c_min =8'd0;
logic rw,g_patt;
logic pass,fail;
integer var_count =8'd0;
```

**Figure 8.16 Toggle coverage details of logic bits in HDL code of MBIST controller.**

**ii) Transition and Sequence Coverage as parts of FSM coverage:** There are 24 possible transitions in FSM of controller which are covered by transition coverage metric shown in Figure8.13 with also 100% state transition coverage. In the Figure8.13, it is shown that total sequence coverage score is 99.88% and all sequences presented in HDL code of controller are covered except one.

| | 0 | 1 | 2 | 3 | 4 | 5 | 6 | 7 | 8 | 9 | 10 | 11 | 12 |
|---|---|---|---|---|---|---|---|---|---|---|---|---|---|
| 0 s_idle | 12/12 | 1/1 | 1/1 | 1/1 | 1/1 | 1/1 | 1/1 | 1/1 | 1/1 | 1/1 | 1/1 | 1/1 | 1/1 |
| 1 wdn0 | 12/12 | 12/12 | 1/1 | 1/1 | 1/1 | 1/1 | 1/1 | 1/1 | 1/1 | 1/1 | 1/1 | 1/1 | 1/1 |
| 2 rup0 | 12/12 | 11/11 | 11/11 | 1/1 | 1/1 | 1/1 | 1/1 | 1/1 | 1/1 | 1/1 | 1/1 | 1/1 | 1/1 |
| 3 wup1 | 13/13 | 12/12 | 10/10 | 10/10 | 1/1 | 1/1 | 1/1 | 1/1 | 1/1 | 1/1 | 1/1 | 1/1 | 1/1 |
| 4 rdn1 | 14/14 | 13/13 | 11/11 | 9/9 | 9/9 | 1/1 | 1/1 | 1/1 | 1/1 | 1/1 | 1/1 | 1/1 | 1/1 |
| 5 wdna0 | 14/14 | 13/13 | 11/11 | 9/9 | 8/8 | 8/8 | 1/1 | 1/1 | 1/1 | 1/1 | 1/1 | 1/1 | 1/1 |
| 6 pause | 14/14 | 13/13 | 11/11 | 9/9 | 8/8 | 7/7 | 7/7 | 1/1 | 1/1 | 1/1 | 1/1 | 1/1 | 1/1 |
| 7 rdn0 | 14/14 | 13/13 | 11/11 | 9/9 | 8/8 | 7/7 | 6/6 | 6/6 | 1/1 | 1/1 | 1/1 | 1/1 | 1/1 |
| 8 wdn1 | 14/14 | 13/13 | 11/11 | 9/9 | 8/8 | 7/7 | 6/6 | 5/5 | 5/5 | 1/1 | 1/1 | 1/1 | 1/1 |
| 9 rup1 | 15/15 | 14/14 | 12/12 | 10/10 | 9/9 | 8/8 | 7/7 | 6/6 | 4/4 | 4/4 | 1/1 | 1/1 | 1/1 |
| 10 wup0 | 15/15 | 14/14 | 12/12 | 10/10 | 9/9 | 8/8 | 7/7 | 6/6 | 4/4 | 3/3 | 3/3 | 1/1 | 1/1 |
| 11 rdna0 | 12/12 | 11/11 | 9/9 | 7/7 | 5/5 | 4/4 | 4/4 | 4/4 | 2/2 | 2/2 | 2/2 | 2/2 | 1/1 |
| 12 s_done | 12/12 | 11/11 | 9/9 | 7/7 | 5/5 | 4/4 | 4/4 | 4/4 | 2/2 | 2/2 | 2/2 | 1/1 | 0/1 |

**Figure 8.17 No. of sequences with their starting and ending states present in controller.**

The first column and first row in the Figure 8.17 shows the states present in MBIST controller. The Figure 8.17 shows the total number of sequences with starting states of sequences while including how many times a particular state became part of sequences. For example, there are 14 sequences started with pause state in which these sequences include the s_idle state and these all sequences have been covered. The dark shaded cell in Figure 8.17 shows the untested or uncovered sequence which start from s_done state and again include itself in the sequence.

**iii) Assertion Coverage:** As shown in Figure8.18, the successful execution of assertions is shown by shaded portion. And total coverage score for the assertion coverage is 100% shown in Figure 8.7with the total functional coverage report of MBIST controller.

```
property done_idle;
@(posedge clk) ((state==done)|=>((state ==s_idle)&&(rst))); endproperty
Ap33:assert property(done_idle);
Cp33:cover property(done_idle);

sequence s1;
state ==wdn0; endsequence
sequence s2;
!(rw||g_patt); endsequence
property f_wdn0;
@(posedge clk) s1 |->s2 ; endproperty
Ap34:assert property(f_wdn0);
Cp34:cover property(f_wdn0);
```

**Figure 8.18Covered assertions shown in gui mode.**

It is a very tedious task to write the test cases for every functional specification of design to cover or verify it and it becomes impractical for large and complex designs. Assertions are included to validate the MBIST controller design by increasing the functional coverage.

# Chapter 9

# Conclusion

The thesis work presents assertion based functional verification of RTL representation of a digital design. As a digital design, a MBIST controller is designed which is based on a memory testing March algorithm. This March algorithm is a little modified March C algorithm which is modified by adding a pause element to test memory data retention faults. In assertion based functional verification, creation of verification plan, for MBIST controller RTL model and the implementation & simulation of the verification plan using system Verilog and Synopsys-VCS® are done. The verification planning is done according to MBIST controller design specification for a complete MBIST architecture and by taking the guide line from the various ABV documentations. In ABV, verification plan includes the MBIST controller design and functional specification, functional coverage goals, code coverage goals and assertions. Assertions are used to check the errors in RTL model of MBIST controller and to provide the functionality coverage. Functional coverage metrics are used to track the level or quality of verification. Most of the functional metrics score approximately reached the planned goal 100 % which is planned in verification plan. That's the indication that all the functionality of design has been verified. Basically in ABV, the designed MBIST controller is verified against intended features. ABV approach helped to make the verification and design process efficient and less time consuming by finding the bugs, exercising the corner cases in the design and by using the directed test cases in a small design. ABV helped to write directed and efficient test cases (25) which are approx 32 %less than the use of maximum possible random test cases (88) for designed MBIST controller with 100% assertion coverage and approximately equal total functional coverage i.e. 97 % approx. In this way ABV helped to fasten the design and verification process with better quality and assurance of correct functionality of MBIST controller after the integration in MBIST architecture.

# Chapter 10

## Recommendation and Future Work

In this work the memory testing algorithm March C is modified to design a MBIST controller which can test SAFs, TFs, AFs, CFs and DRFs in SRAM memories. But there are various new memories types like DDR-4 and these memories have different new faults so the algorithm can be modified to test those new memory faults.

Here, the assertion based functional verification of designed MBIST controller is done so as a future work its integration in complete MBIST architecture is required to test the single port SRAM memories.

The approach followed in functional verification of MBIST controller can be used to verify the all MBIST components and then it will ensures the correct functionality of complete MBIST architecture.

Lastly, for the verification of complete MBIST module is recommended using Verification Methodology Manual (VMM) or Universal Verification Methodology (UVM) verification methodologies.

The use of System Verilog HDVL is suggested for the design and verification purpose which added many advantages to complete design flow.

# Appendix A

## A.1 Source code and Assertions

```
Module mbist_cont#(parameter sign_size =32, c_size =8, size=4)
(input t_mode, clk, rst,match, [1:sign_size] signature,
[1:sign_size]mem_out,    output done,pass,fail);

//Defining design output and inputs in generalized form by
defining parameters.

enum logic [3:0]
{s_idle,wdn0,rup0,wup1,rdn1,wdna0,pause,rdn0,wdn1,rup1,wup0,rd
na0,s_done} state;

// The variable state is defined as enumerated type.

Logic done,pass,fail, en;

logic [1:sign_size] signature;//the size of signature is
                              defined as generalized.
logic [c_size-1:0] c_max = 8'd255;

logic [c_size-1:0] count;

logic [3:0] state;

logic [1:sign_size]mem_out;

logic match ;

logic [c_size-1:0] c_min =8'd0;

logic rw,g_patt;

logic pass,fail;

integer var_count =8'd0;//counter's count is define as the
                         integer data type.


always_ff@(posedge clk or posedge rst)
begin
```

```
   if(rst) begin //Defining signals in idle state of controller
                                    when reset signal is asserted.

state<=s_idle; en<=1'b0;done<=0;fail<=0;pass<=1;

 count<=8'd0;g_patt<=1;rw<=1;//high value of rw signal enables
                                   the read operation over memory.
end

else

case (state)

   s_idle: if(~t_mode) begin//Again controller will be in idle
                       state if t_mode signal is not asserted.
state<=s_idle;
end

else begin

//controller would be in the next state or march state only if
reset signal is de-asserted and t_mode signal is asserted.

     en<=1'b1;
     state<=wdn0; count <= c_max;
     rw<=0;g_patt<=0;
     end

    wdn0:begin//Now controller is in state wdn0.In this state,
              controller controls the generation of march test
              0 pattern for down marching of memory addresses.

count<= count-1; if(count ==c_min)

begin state =rup0;rw<=1;
count<= c_min;
end

else state =wdn0;
end

rup0:begin
count<=count+1;if(match)
begin fail<=0; pass<=1'b1;end

 else begin pass<=0;fail<=1'b1;end //High value of fail signal
```

```
                    indicates the presence of fault in memory.


if(count == c_max) begin
                        state= wup1;g_patt<=1;rw<=0;
                        count<=c_min;
end
else state =rup0;
end


wup1:begin
count<=count+1; if(count==c_max)begin
state=rdn1;rw<=1;
     count<=c_max;

end
else  state= wup1;
end



rdn1:begin
count<=count-1;if(match)begin
pass<=1'b1;fail<=0;
end

else begin
fail<=1'b1;pass<=0;
end

if(count==c_min) begin
                    state =wdna0;g_patt<=0;
                     rw<=0;//Low value of rw signal enable the
                             write operation on memory instead
                                            of read operation.
                    count<=c_max;
                    end

else  state = rdn1;
end


 wdna0:begin
count<=count-1;
```

```
                if(count==c_min) begin
                state=pause;count<=c_min;
                end

else state = wdna0;
end


pause: begin
count<=count+1;
if (count==c_max)begin
state=rdn0;rw<=1;count<=c_max;
end

else state=pause;
end


rdn0:begin
count<=count-1;
if(match) begin
pass<=1'b1;fail<=1'b0;
end

else begin
fail<=1'b1;pass<=1'b0;
end

if(count==c_min)begin
                    state=wdn1;rw<=0; g_patt<=1;count<=c_max;
                    end

else state=rdn0;
end

wdn1:begin
count<=count-1;
if(count==c_min)begin
count<=c_min;rw<=1;state=rup1;
end

else state= wdn1;
end

rup1:begin
count<=count+1; if(match)begin
```

```
   pass<=1'b1;fail<=1'b0; //if signal match is high, then only
                   pass signals must high and fail signal low.

end

else begin
fail<=1'b1; pass<=1'b0;
end

if(count==c_max)begin
state=wup0;g_patt<=0; rw<=0;count<=c_min;
end

else state=rup1;
end


wup0:begin
count<=count+1;
if(count==c_max)begin
count<=c_max;rw<=1;state=rdna0;
end

else state=wup0;
end

rdna0:begin
count<=count-1;
if(match) begin
pass<=1'b1;fail<=1'b0;
end

else begin
fail<=1'b1; pass<=1'b0;
end

if(count==c_min) begin
state=s_done;done<=1'b1;en<=0;
end

else state = rdna0;
end


s_done: begin
```

```
//After completion of test or traversing all march states
controller state will be final state as Done state.

   if(t_mode) state=s_done; //In state s_done, the signal done
                                          will be having value '1'.
else state=s_idle;
end


endcase

end



//Below lines of code are embedded assertions in the

sourcecode of MBIST controller.


sequence seq1;

(state==wdn0)&&(count!==c_min);endsequence

property loop_wdn0;

@(posedgeclk) disable iff(state!==wdn0) seq1|-> (state==wdn0);

endproperty

Ap_loop_wdn0:assert property(loop_wdn0);

Cp_loop_wdn0:cover property(loop_wdn0);


sequence seq2;

(state==pause)&&(count!==c_max);endsequence

propertyloop_pause;

@(posedgeclk) disable iff(state!==pause) seq2|->

(state==pause); endproperty

Ap_loop_pause:assert property(loop_pause);

Cp_loop_pause:cover property(loop_pause);


sequence seq4;

(state==done)&&(t_mode);endsequence
```

```
Propertyloop_done;
@(posedgeclk) disable iff(state!==done) seq4|-> (state==done);
endproperty
Ap_loop_done:assert property(loop_done);
Cp_loop_done:cover property(loop_done);
sequence seq3;
(state==rdna0)&&(count!==c_min);endsequence
property loop_rdna0;
@(posedgeclk) disable iff(state!==rdna0) seq3|->
(state==rdna0); endproperty
Ap_loop_rdna0:assert property(loop_rdna0);
Cp_loop_rdna0:cover property(loop_rdna0);

sequence check_rup0;
(state==rup0);endsequence
sequencecheck_idle;
((rst==1)&&(state==s_idle));endsequence
property rup0_idle;
@(posedgeclk) check_rup0|=>check_idle; endproperty
Ap_rup0_idle:assert property(rup0_idle);
Cp_rup0_idle:cover property(rup0_idle);

property wdn0_idle;
@(posedgeclk)((state==wdn0)|=>(rst==1)&&(state==s_idle));
endproperty
Ap_wdn0_idle:assert property(wdn0_idle);
Cp_wdn0_idle:cover property(wdn0_idle);

sequence check_wup1;
state==wup1;endsequence
sequence check2_idle;
```

```
((rst==1)&&(state==s_idle));endsequence
property wup1_idle;
@(posedgeclk) check_wup1|=>check2_idle; endproperty
Ap_wup1_idle:assert property(wup1_idle);
Cp_wup1_idle:cover property(wup1_idle);

sequence check_rdn1;
state==rdn1;endsequence
sequence check3_idle;
((rst==1)&&(state==s_idle));endsequence
property rdn1_idle;
@(posedgeclk)check_rdn1|=>check3_idle; endproperty
Ap_rdn1_idle:assert property(rdn1_idle);
Cp_rdn1_idle:cover property(rdn1_idle);

sequence check_wdna0;
state==wdna0;endsequence
sequence check4_idle;
((rst==1)&&(state==s_idle));endsequence
property wdna0_idle;
@(posedgeclk) check_wdna0|=>check4_idle; endproperty
Ap_wdna0_idle:assert property(wdna0_idle);
Cp_wdna0_idle:cover property(wdna0_idle);

sequence check_rdn0;
state==rdn0;endsequence
sequence check5_idle;
((rst==1)&&(state==s_idle));endsequence
property rdn0_idle;
@(posedgeclk) check_rdn0|=>check5_idle; endproperty
Ap_rdn0_idle:assert property(rdn0_idle);
```

```
Cp_rdn0_idle:cover property(rdn0_idle);

sequence check_wdn1;
state==wdn1;endsequence
sequence check6_idle;
((rst==1)&&(state==s_idle));endsequence
property wdn1_idle;
@(posedgeclk)check_wdn1|=>check6_idle; endproperty
Ap_wdn1_idle:assert property(wdn1_idle);
Cp_wdn1_idle:cover property(wdn1_idle);

sequence check_rup1;
state==rup1;endsequence
sequence check7_idle;
((rst==1)&&(state==s_idle));endsequence
property rup1_idle;
@(posedgeclk) check_rup1|=>check6_idle; endproperty
Ap_rup1_idle:assert property(rup1_idle);
Cp_rup1_idle:cover property(rup1_idle);

sequence check_wup0;
state==wup0;endsequence
sequence check8_idle;
((rst==1)&&(state==s_idle));endsequence
property wup0_idle;
@(posedgeclk)  check_wup0|=>check8_idle; endproperty
Ap_wup0_idle:assert property(wup0_idle);
Cp_wup0_idle:cover property(wup0_idle);

sequence check_rdna0;
state==rdna0;endsequence
```

```
sequence check9_idle;
((rst==1)&&(state==s_idle));endsequence
property rdna0_idle;
@(posedgeclk)  check_rdna0|=>check9_idle; endproperty
Ap_rdna0_idle:assert property(rdna0_idle);
Cp_rdna0_idle:cover property(rdna0_idle);

Sequencecheck_s_done;
state==s_done;endsequence
sequence check10_idle;
((rst==1)&&(state==s_idle));endsequence
Propertys_done_idle;
@(posedgeclk) check_s_done|=>check10_idle; endproperty
Ap_s_done_idle:assert property(s_done_idle);
Cp_s_done_idle:cover property(s_done_idle);

sequence stat_wdn0;
(state==wdn0) && (count==c_min); endsequence
property rup0_tran;
@(posedgeclk) stat_wdn0|-> ##1 state==rup0; endproperty
Ap13:assert property(rup0_tran)
$display ("rup0 state transition is legal");
else
$warning ("rup0 state transition is illegal");
Cp13:cover property(rup0_tran);

/*property t_mode_idle;
@(posedgeclk) disable iff (t_mode) (state==s_idle);
endproperty
Ap_t_mode_idle: assert property(t_mode_idle);
cp_t_mode_idle:cover property(t_mode_idle); defined as ap9*/
```

```
propertystate_stable;
@(posedgeclk) rst |-> $stable(state); endproperty
Ap6: assert property (state_stable)
$display("High rst no state transition");
else
$warning ("state transition is not stable with rst");
Cp6:cover property (state_stable);

Propertyfsm_rst;
@(posedgeclk) rst |-> state ==s_idle;
endproperty
Ap1: assert property (fsm_rst)
 $display ("property is successed\n");
else
$info ("property is failed\n");
Cp1: cover property(fsm_rst);

Sequenceen_t;
 (t_mode&& en); endsequence
sequence s_wdn0;
 (state ==wdn0); endsequence
property en_wdn0;
@(posedgeclk)first_match(s_wdn0)|->en_t; endproperty
Ap2:assert property(en_wdn0)
  $display ("enable rose at right time");
else
 $warning ("enable is not at right time");
Cp2:cover property(en_wdn0);

sequence en_t1;
```

```
 (t_mode&& en); endsequence
sequence s_rup0;
 (state ==rup0); endsequence
property en_rup0;
@(posedgeclk)first_match(s_rup0)|->en_t1; endproperty
Ap3:assert property(en_rup0);
Cp3:cover property(en_rup0);

sequence en_t2;
 (t_mode&& en); endsequence
sequence s_wup1;
 (state ==wup1); endsequence
property en_wup1;
@(posedgeclk)first_match(s_wup1)|->en_t2; endproperty
Ap4:assert property(en_wup1);
Cp4:cover property(en_wup1);

Propertycheck_en ;
@(posedgeclk) disable iff(!t_mode) (t_mode)|->en; endproperty
Ap5:assert property(check_en)
 $display("enable rise is ok");
else
 $warning("enable gets delay");
Cp5:cover property (check_en);

property pause_rdn0;
@(posedgeclk) (state==pause) |-> ##256 (state ==rdn0);
endproperty
Ap7: assert property(pause_rdn0);
Cp7:cover property(pause_rdn0);
```

```
propertyrose_t_mode;
@(posedgeclk) disable iff(rst) (t_mode); endproperty
Ap8: assert property (rose_t_mode)
 $display("t_mode and rst are synchronized");
else
 $error("t_mode and rst are not synchronized");
Cp8:cover property (rose_t_mode);

propertyfsm_idle;
@(posedgeclk) disable iff (rst==1)!t_mode |-> state ==s_idle;
endproperty
Ap9:assert property (fsm_idle);
Cp9:cover property(fsm_idle);



Sequencestate_pause;
(state==pause) && (count==c_max); endsequence
property rdn0_tran;
@(posedgeclk) state_pause|=> (state==rdn0); endproperty
Ap10:assert property(rdn0_tran)
$display ("rdn0 state transition is legal");
else
$warning ("rdn0 state transition is illegal");
Cp10:cover property(rdn0_tran);

Sequencestate_idle;
(state==s_idle) && (count==c_min); endsequence
property wdn0_tran;
@(posedgeclk) state_idle|-> ##1 state==wdn0; endproperty
Ap12:assert property(wdn0_tran)
$display ("wdn0 state transition is legal");
```

```
else
$warning ("wdn0 state transition is illegal");
Cp12:cover property(wdn0_tran);

sequence state_rup0;
(state==rup0) && (count==c_max); endsequence
property wup1_tran;
@(posedgeclk) (state_rup0)|-> ##1 state==wup1; endproperty
Ap14:assert property(wup1_tran)
$display ("wup1 state transition is legal");
else
$warning ("wup1 state transition is illegal");
Cp14:cover property(wup1_tran);

Propertyrst_t_mode;
@(posedgeclk) !(rst&&t_mode); endproperty
Ap11: assert property(rst_t_mode)
     $display ("input constraints is ok\n");
else
     $error ("fatal is occured\n");
Cp11:cover property(rst_t_mode);

Propertymatch_pass;
@(posedgeclk) match |-> pass; endproperty
Ap15:assert property (match_pass);
Cp15:cover property (match_pass);

Propertymatch_fail;
@(posedgeclk) match |-> !(fail && pass);endproperty
Ap16:assert property(match_fail);
Cp16:cover property(match_fail);
```

```
propertyp_f;
@(posedgeclk) t_mode |-> !(pass && fail); endproperty
Ap17: assert property(p_f);
Cp17: cover property(p_f);




property wdna0_pause;
@(posedgeclk) ((state ==wdna0)|=>((state ==pause)
&&(count==c_min)));endproperty
Ap18:assert property(wdna0_pause);
Cp18: cover property (wdna0_pause);

property idl_wdn0;
@(posedgeclk) ((state ==s_idle)|=>((state ==wdn0)
&&(t_mode)));endproperty
Ap19:assert property(idl_wdn0);
Cp19: cover property (idl_wdn0);

property wdn0_rup0;
@(posedgeclk) ((state ==wdn0)|=>((state ==rup0)
&&(count==c_min)));endproperty
Ap20:assert property(wdn0_rup0);
Cp20: cover property (wdn0_rup0);

property rup0_wup1;
@(posedgeclk) ((state ==rup0)|=>((state ==wup1)
&&(count==c_min)));endproperty
Ap21:assert property(rup0_wup1);
Cp21: cover property (rup0_wup1);
```

```
property wup1_rdn1;
@(posedgeclk) ((state ==wup1)|=>((state ==rdn1)
&&(count==c_max)));endproperty
Ap22:assert property(wup1_rdn1);
Cp22: cover property (wup1_rdn1);

property rdn1_wdna0;
@(posedgeclk) ((state ==rdn1)|=>((state ==wdna0)
&&(count==c_max)));endproperty
Ap23:assert property(rdn1_wdna0);
Cp23: cover property (rdn1_wdna0);



Propertyp_pause;
@(posedgeclk)((state ==pause)|=>((state ==pause) && !(count
==c_max))); endproperty
Ap24: assert property(p_pause);
Cp24:cover property (p_pause);

property paus_rdn0;
@(posedgeclk) ((state ==pause)|=>((state
==rdn0)&&(count==c_max))); endproperty
Ap25: assert property(paus_rdn0);
Cp25:cover property(paus_rdn0);

property rdn0_wdn1;
@(posedgeclk) ((state==rdn0)|=>((state
==wdn1)&&(count==c_max))); endproperty
Ap26:assert property(rdn0_wdn1);
Cp26:cover property (rdn0_wdn1);
```

```
property wdn1_rup1;
@(posedge
clk)((state==wdn1)|=>((state==rup1)&&(count==c_min)));endprope
rty
Ap27:assert property(wdn1_rup1);
Cp27:cover property (wdn1_rup1);

property rup1_wup0;
@(posedgeclk)((state==rup1)|=>
((state==wup0)&&(count==c_min))); endproperty
Ap28:assert property(rup1_wup0);
Cp28:cover property (rup1_wup0);

property wup0_rdna0;
@(posedgeclk)
((state==wup0)|=>((state==rdna0)&&(count==c_max)));endproperty
Ap29:assert property (wup0_rdna0);
Cp29:cover property(wup0_rdna0);

property r_rdna0;
@(posedgeclk)
((state==rdna0)|=>((state==rdna0)&&!(count==c_min)));
endproperty
Ap30:assert property(r_rdna0);
Cp30:cover property(r_rdna0);

property rdna0_done;
@(posedgeclk)
((state==rdna0)|=>((state==s_done)&&(count==c_max)));
endproperty
```

```
Ap31:assert property(rdna0_done);
Cp31:cover property(rdna0_done);

Propertyd_done;
@(posedgeclk)
((state==done)|=>((state==done)&&(!rst)));endproperty
Ap32:assert property(d_done);
Cp32:cover property(d_done);

propertydone_idle;
@(posedgeclk) ((state==done)|=>((state ==s_idle)&&(rst)));
endproperty
Ap33:assert property(done_idle);
Cp33:cover property(done_idle);

sequence s1;
state ==wdn0; endsequence
sequence s2;
!(rw||g_patt); endsequence
property f_wdn0;
@(posedgeclk) s1 |->s2 ; endproperty
Ap34:assert property(f_wdn0);
Cp34:cover property(f_wdn0);

Propertycheck_idle_en;
@(posedgeclk)disableiff(state!==s_idle) ##1 en==0; endproperty
Ap_check_idle_en:assert property(check_idle_en);
Cp_check_idle_en:cover property(check_idle_en);

Propertycheck_en_value;
@(posedgeclk)disable iff(state==s_idle)en==1;endproperty
```

```
Ap_check_en_value:assert property(check_en_value);
Cp_check_en_value:cover property(check_en_value);

Propertycheck_pause;
@(posedgeclk)disable iff(state!==pause)
(en==1)|-> $stable(rw);endproperty
Ap_check_pause:assert property(check_pause);
Cp_check_pause:cover property(check_pause);

Propertycheck_done;
@(posedgeclk)disable iff(state!==s_done)
##1(done==1)|->(en==0);endproperty
Ap_check_done:assert property(check_done);
Cp_check_done:cover property(check_done);
Endmodule
```

### A.2 All Assertions and their explanation

**Table A.1 Name of all used assertions and their explanations.**

| S.No. | Name of Assertion | Purpose of Used |
|---|---|---|
| 1. | Ap_loop_wdno | To Check the address counter value from maximum to minimum address while writing 0 pattern only in state wdn0. |
| 2. | Ap_loop_pause | To check counter values from its min to max count values in pause state. |
| 3. | Ap_loop_rdna0 | To check address counter value in rdna0 state only. |
| 4. | Ap_loop_done | To verify that Done is the stable state of controller until t_mode signal is high. |
| 5. | Ap_rup0_idle | To check the state transition from rup0 to idle when rst is asserted. |
| 6. | Ap_wdn0_idle | To check the transition from wdn0 to idle when rst is asserted. |
| 7. | Ap_wup1_idle | To check the transition from wup1 to idle when rst is asserted. |
| 8. | Ap_rdn1_idle | To check the transition from rdn1 to idle when rst is asserted. |
| 9. | Ap_wdna0_idle | To check the transition from wdna0 to idle when rst is asserted |
| 10. | Ap_rdn0_idle | To check the transition from rdn0 to idle when rst is asserted. |
| 11. | Ap_wdn1_idle | To check the transition from wdn1 to idle when rst is asserted. |
| 12. | Ap_rdn1_idle | To check the transition from rdn1 to idle when rst is asserted. |
| 13. | Ap_wup0_idle | To check the transition from wup0 to idle when rst is asserted. |
| 14. | Ap_rdna0_idle | To check the transition from rdna0 to idle when rst is asserted. |
| 15. | Ap_s_done_idle | To check the transition from done to idle when rst is asserted. |
| 16. | Ap13 | To check the state transition from state wdn0 to rup0. |
| 17. | Ap6 | To check whether state transition is stable or not when rst is asserted. |
| 18. | Ap1 | To check whether controller goes to idle state or not with in 1 clock cycle after asserting rst signal. |
| 19. | Ap2 | Checks the desired values of t_mode and en signals in wdn0 state. |
| 20. | Ap3 | First checks the state rup0 and then checks the desired values of t_mode and en signals. |
| 21. | Ap4 | First matches the state wup1 and then checks the t_mode and en signals. |
| 22. | Ap5 | Checks the synchronization between t_mode and en signals. |
| 23. | Ap7 | To Verify whether the test operation in pause stae would be hold up to a fixed time or not and then controller goes to next state rdn0. |
| 24. | Ap8 | To check the synchronization between t_mode and rst signals. |
| 25. | Ap9 | To verify that controller must be idle until t_mode is asserted even if rst is already deasserted. |
| 26. | Ap10 | To check the legal state transition from pause to rdn0. |
| 27. | Ap12 | To checks whether controller changes its state from state idle to wdn0 with minimum address counter value. |
| 28. | Ap14 | To check the legal state transition from rup0 to wup1. |
| 29. | Ap11 | To check the set constraints on the t_mode and rst signals at every clock pulse. |
| 30. | Ap15 | To checks the synchronization and values of signal match and pass. |
| 31. | Ap16 | To check the synchronization and values of signals match and fail. |
| 32. | Ap17 | To verify that signal fail and pass cannot have same values if t_mode is high. |
| 33. | Ap18 | To check the state transition from wdna0 to pause only if address count value is minimum. |
| 34. | Ap19 | To verify that state transition from idle to wdn0 only if t_mode is high. |
| 35. | Ap20 | To check the legal state transition from wdn0 to rup0. |
| 36. | Ap21 | To check the legal state transition from rup0 to wup1. |
| 37. | Ap22 | To verify that state rdn1 has initial address counter value is maximum count value after transition from state wup1. |
| 38. | Ap23 | To check whether state transition from rdn1 to wdna0 is legal or not. |
| 39. | Ap24 | First matches the pause state and then checks address counter count values. |
| 40. | Ap25 | First matches the pause state and checks the counter values as maximum from state transition from pause to rdn0. |
| 41. | Ap26 | To check the legal state transition from rdn0 to wdn1. |
| 42. | Ap27 | To check the legal state transition from wdn1 to rup1. |
| 43. | Ap28 | To check the legal state transition from rup1 to wup0. |
| 44. | Ap29 | To check the legal state transition from wup0 to rdna0. |
| 45. | Ap30 | To verify the correct order and count values of counter during state rdna0. |
| 46. | Ap31 | To check legal state transition to state done. |
| 47. | Ap32 | To verify, if present state is done than it will be remains same until rst becomes high. |
| 48. | Ap33 | To check the state transition from done to idle if rst=1. |
| 49. | Ap34 | To check the values of signals rw and g_patt in state wdn0. |
| 50. | Ap_check_idle_en | To check the en=0, only when state is idle. |
| 51. | Ap_check_en_value | To check the en=1, if state is not idle. |
| 52. | Ap_check_pause | To check that the signal rw will be state in pause state. |
| 53. | Ap_check_done | To verify if done=1, in the next clk cycle signal en must be 0. |

# Appendix B

**Detailed Assertions and Cover Properties Report**

Assertions

Assertions by Category

| | ASSERT | PROPERTIES | SEQUENCES |
|---|---|---|---|
| Total | 53 | 53 | 0 |
| Category 0 | 53 | 53 | 0 |

------------------------------------------------------------------------------

Assertions by Severity

| | ASSERT | PROPERTIES | SEQUENCES |
|---|---|---|---|
| Total | 53 | 53 | 0 |
| Severity 0 | 53 | 53 | 0 |

------------------------------------------------------------------------------

**Detail Report for Assertions**

| ASSERTIONS | ATTEMPTS | REAL SUCCESSES | FAILURES | INCOMPLETE |
|---|---|---|---|---|
| tb_bist_cont.uut.Ap1 | 21000 | 27 | 0 | 0 |
| tb_bist_cont.uut.Ap | 21000 | 7 | 0 | 0 |
| tb_bist_cont.uut.Ap11 | 21000 | 20975 | 25 | 0 |
| tb_bist_cont.uut.Ap12 | 21000 | 13 | 32 | 0 |
| tb_bist_cont.uut.Ap13 | 21000 | 12 | 0 | 0 |
| tb_bist_cont.uut.Ap14 | 21000 | 11 | 0 | 0 |
| tb_bist_cont.uut.Ap15 | 21000 | 20954 | 19 | 0 |
| tb_bist_cont.uut.Ap16 | 21000 | 20973 | 0 | 0 |
| tb_bist_cont.uut.Ap17 | 21000 | 20991 | 0 | 0 |
| tb_bist_cont.uut.Ap18 | 21000 | 8 | 2064 | 0 |
| tb_bist_cont.uut.Ap19 | 21000 | 13 | 32 | 0 |
| tb_bist_cont.uut.Ap2 | 21000 | 3322 | 0 | 0 |
| tb_bist_cont.uut.Ap20 | 21000 | 12 | 3310 | 0 |
| tb_bist_cont.uut.Ap21 | 21000 | 11 | 2821 | 0 |
| tb_bist_cont.uut.Ap22 | 21000 | 10 | 2564 | 0 |
| tb_bist_cont.uut.Ap23 | 21000 | 9 | 2309 | 0 |
| tb_bist_cont.uut.Ap24 | 21000 | 1933 | 15 | 0 |
| tb_bist_cont.uut.Ap25 | 21000 | 7 | 1941 | 0 |
| tb_bist_cont.uut.Ap26 | 21000 | 6 | 1653 | 0 |
| tb_bist_cont.uut.Ap27 | 21000 | 5 | 1294 | 0 |
| tb_bist_cont.uut.Ap28 | 21000 | 4 | 1054 | 0 |

```
tb_bist_cont.uut.Ap29       21000    3              781      0
tb_bist_cont.uut.Ap3        21000    2832           0        0
tb_bist_cont.uut.Ap30       21000    526            5        0
tb_bist_cont.uut.Ap31       21000    2              529      0
tb_bist_cont.uut.Ap32       21000    18             27       0
tb_bist_cont.uut.Ap33       21000    14             31       0
tb_bist_cont.uut.Ap34       21000    3322           0        0
tb_bist_cont.uut.Ap4        21000    2574           0        0
tb_bist_cont.uut.Ap5        21000    20397          594      0
tb_bist_cont.uut.Ap6        21000    14             13       0
tb_bist_cont.uut.Ap7        21000    1659           289      0
tb_bist_cont.uut.Ap8        21000    20955          5        0
tb_bist_cont.uut.Ap9        21000    5              0        0
tb_bist_Ap_check_done21000    554            0        1
tb_bist_Ap_check_en21000    20386          556      0
tb_bist_Ap_check_idle_en21000    19             0        0
tb_bist_Ap_check_pause21000    1940           0        0
tb_bist_Ap_loop_done21000    24             0        0
tb_bist_Ap_loop_pause 21000    1940           0        0
tb_bist_Ap_loop_rdna0 21000    528            0        0
tb_bist_Ap_loop_wdn0  21000    3309           0        0
tb_bist_Ap_rdn0_idle  21000    1              1658     0
tb_bist_Ap_rdn1_idle  21000    1              2317     0
tb_bist_Ap_rdna0_idle 21000    1              530      0
tb_bist_Ap_rup0_idle  21000    1              2831     0
tb_bist_Ap_rup1_idle  21000    1              1057     0
tb_bist_Ap_s_done_idle21000    1              555      1
tb_bist_Ap_wdn0_idle  21000    1              3321     0
tb_bist_Ap_wdn1_idle21000    1              1298     0
tb_bist_Ap_wdna0_ide  21000    1              2071     0
tb_bist_Ap_wup0_idLe21000    1              783      0
tb_bist_Ap_wup1_idle 21000    1              2573     0
```

------------------------------------------------------------------------------

**Detail Report for Cover Properties**

```
COVER PROPERTIES     ATTEMPTS MATCHES INCOMPLETE
tb_bist_cont.uut.Cp1           21000    27      0
tb_bist_cont.uut.Cp10             21000    7       0
tb_bist_cont.uut.Cp11             21000    20975   0
tb_bist_cont.uut.Cp12             21000    13      0
tb_bist_cont.uut.Cp13             21000    12      0
tb_bist_cont.uut.Cp14             21000    11      0
tb_bist_cont.uut.Cp15             21000    20954   0
tb_bist_cont.uut.Cp16             21000    20973   0
tb_bist_cont.uut.Cp17             21000    20991   0
tb_bist_cont.uut.Cp18             21000    8       0
tb_bist_cont.uut.Cp19             21000    13      0
```

| | | | |
|---|---|---|---|
| tb_bist_cont.uut.Cp2 | 21000 | 3322 | 0 |
| tb_bist_cont.uut.Cp20 | 21000 | 12 | 0 |
| tb_bist_cont.uut.Cp21 | 21000 | 11 | 0 |
| tb_bist_cont.uut.Cp22 | 21000 | 10 | 0 |
| tb_bist_cont.uut.Cp23 | 21000 | 9 | 0 |
| tb_bist_cont.uut.Cp24 | 21000 | 1933 | 0 |
| tb_bist_cont.uut.Cp25 | 21000 | 7 | 0 |
| tb_bist_cont.uut.Cp26 | 21000 | 6 | 0 |
| tb_bist_cont.uut.Cp27 | 21000 | 5 | 0 |
| tb_bist_cont.uut.Cp28 | 21000 | 4 | 0 |
| tb_bist_cont.uut.Cp29 | 21000 | 3 | 0 |
| tb_bist_cont.uut.Cp3 | 21000 | 2832 | 0 |
| tb_bist_cont.uut.Cp30 | 21000 | 526 | 0 |
| tb_bist_cont.uut.Cp31 | 21000 | 2 | 0 |
| tb_bist_cont.uut.Cp32 | 21000 | 18 | 0 |
| tb_bist_cont.uut.Cp33 | 21000 | 14 | 0 |
| tb_bist_cont.uut.Cp34 | 21000 | 3322 | 0 |
| tb_bist_cont.uut.Cp4 | 21000 | 2574 | 0 |
| tb_bist_cont.uut.Cp5 | 21000 | 20397 | 0 |
| tb_bist_cont.uut.Cp6 | 21000 | 14 | 0 |
| tb_bist_cont.uut.Cp7 | 21000 | 1659 | 0 |
| tb_bist_cont.uut.Cp8 | 21000 | 20955 | 0 |
| tb_bist_cont.uut.Cp9 | 21000 | 5 | 0 |
| tb_bist_cont.uutCp_check_done | 21000 | 554 | 1 |
| tb_bist_cont.uutCp_check_en_value | 21000 | 20386 | 0 |
| tb_bist_cont.uutCp_check_idle_en | 21000 | 19 | 0 |
| tb_bist_cont.uutCp_check_pause | 21000 | 1940 | 0 |
| tb_bist_cont.uutCp_loop_done | 21000 | 24 | 0 |
| tb_bist_cont.uutCp_loop_pause | 21000 | 1940 | 0 |
| tb_bist_cont.uutCp_loop_rdna0 | 21000 | 528 | 0 |
| tb_bist_cont.uutCp_loop_wdn0 | 21000 | 3309 | 0 |
| tb_bist_cont.uutCp_rdn0_idle | 21000 | 1 | 0 |
| tb_bist_cont.uutCp_rdn1_idle | 21000 | 1 | 0 |
| tb_bist_cont.uutCp_rdna0_idle | 21000 | 1 | 0 |
| tb_bist_cont.uutCp_rup0_idle | 21000 | 1 | 0 |
| tb_bist_cont.uutCp_rup1_idle | 21000 | 1 | 0 |
| tb_bist_cont.uutCp_s_done_idle | 21000 | 1 | 1 |
| tb_bist_cont.uutCp_wdn0_idle | 21000 | 1 | 0 |
| tb_bist_cont.uutCp_wdn1_idle | 21000 | 1 | 0 |
| tb_bist_cont.uutCp_wdna0_idle | 21000 | 1 | 0 |
| tb_bist_cont.uutCp_wup0_idle | 21000 | 1 | 0 |
| tb_bist_cont.uutCp_wup1_idle | 21000 | 1 | 0 |